\documentclass{article}
\usepackage[utf8]{inputenc} % allow utf-8 input
\usepackage[T1]{fontenc}    % use 8-bit T1 fonts
\usepackage{lmodern}        % scalable fonts for microtype

\usepackage{hyperref}       % hyperlinks
\usepackage{url}            % simple URL typesetting
\usepackage{booktabs}       % professional-quality tables
\usepackage{amsfonts}       % blackboard math symbols
\usepackage{nicefrac}       % compact symbols for 1/2, etc.
\usepackage{microtype}      % microtypography
\usepackage[square,numbers]{natbib}
\usepackage{doi}
\usepackage[dvipsnames]{xcolor}
\usepackage[russian,english]{babel}
\usepackage{epstopdf} %converting to PDF

\usepackage{graphicx}
\usepackage{amsmath,amssymb}
\usepackage{dsfont}
\usepackage{authblk}

\usepackage{fancyhdr}
\parskip=\medskipamount

\newcommand{\fig}[1]{Fig.~\ref{#1}}

\newcommand{\be}{\begin{equation}}
\newcommand{\ee}{\end{equation}}

\usepackage{float}
\usepackage[font=small]{caption}

\date{}
\title{INTENSE: Information-theoretic discovery of behavioral selectivity of neurons in calcium imaging data}
\hypersetup{pdftitle={INTENSE: Information-theoretic discovery of behavioral selectivity of neurons in calcium imaging data}}

\author{Nikita Pospelov$^{1,*}$, Olga Ivashkina$^1$, Viktor Plusnin$^1$, Olga Rogozhnikova$^1$, Anna
Ivanova$^1$, Vladimir Sotskov$^2$, Ksenia Toropova$^1$, Vladik Avetisov$^3$ and Konstantin Anokhin$^1$}

\affil{
$^1$Institute for Advanced Brain Studies of the Moscow State University, Moscow,
Russia}
\affil{
$^2$Center for Interdisciplinary Research in Biology, Coll\`ege de France, Paris, France}
\affil{
$^3$Semenov Institute of Chemical Physics, Russian Academy of Sciences, Moscow, Russia}
\affil{
$^*$Correspondence: Nikita Pospelov (\href{mailto:pospelov.na14@physics.msu.ru}{pospelov.na14@physics.msu.ru})}

\usepackage{pdfpages}
\begin{document}
\maketitle

\begin{abstract}
Neurons in freely behaving animals exhibit selectivity for diverse variables. Calcium imaging makes this selectivity hard to read out: behavioral variables covary, and fluorescence is slow and autocorrelated. We introduce INTENSE (INformation-Theoretic Evaluation of Neuronal SElectivity), an open-source framework that screens every neuron against all measured variables on a common statistical scale. INTENSE applies mutual information directly to fluorescence to detect relationships missed by linear correlation, optimizes neural--behavioral delays, performs autocorrelation-preserving circular-shift tests and examines whether apparent mixed selectivity reflects behavioral covariance. In ground-truth simulations, it reduced false associations caused by temporal structure and recovered more true associations than correlation-based screening. In CA1 recordings from 16 freely exploring mice, INTENSE identified neurons selective for position, head direction, object interaction and locomotor state. Accounting for correlated variables reclassified a large fraction of apparently mixed-selective neuron--session observations. INTENSE turns neural and behavioral recordings into testable maps of neuronal selectivity.
\end{abstract}

A central question in systems neuroscience is which aspects of an animal's ongoing behavior are reflected in the activity of individual neurons, and how to answer it when behavior is high-dimensional, correlated and structured in time \cite{shvyrkov1985psychophysiological, DayanAbbott2001, krakauer2017neuroscience, stringer2019spontaneous, buzsaki2020brain}. This problem is not only statistical but also conceptual. Most analysis pipelines begin with experimenter-defined categories such as place cells \cite{o1971hippocampus}, speed cells \cite{kropff2015speed}, or task-relevant variables \cite{mix3}, implicitly assuming that the relevant ontology of neural selectivity is already known. However, in natural behavior, neuronal activity can relate to measured variables that are not canonical, not task-defined, and not obvious a priori \cite{Musall2019, stringer2019spontaneous, datta2019computational}. A useful neuronal selectivity analysis should therefore be exploratory rather than pre-defined.

Neuronal selectivity is dynamically shaped by experience. In the hippocampus, place cells can rapidly form spatial receptive fields \cite{Sotskov2022}, and in many settings these maps can remain stable over weeks \cite{Thompson1990}. Experience also shapes how variables are combined: many neurons show mixed selectivity \cite{TyeMiller2024}, responding to conjunctions of events, behaviors or cognitive states rather than to a single variable. Prefrontal neurons, for example, develop mixed selectivity for task-relevant variables through learning \cite{mix3, Mante2013, Dang2021}. Even canonical examples of functionally selective neurons can encode multiple variables: place cells often reflect not only location, but also time, rewards, and trajectories \cite{Zheng2021, MacDonald2011}. This principle extends to sensory processing: neurons in the primary visual cortex and retrosplenial cortex can be modulated by both currently perceived and previously experienced visual content \cite{Makino2015}. Because a single neuron can respond to several variables, its apparent selectivity for any one of them may be carried by another variable that changes with it. Selectivity therefore has to be attributed, not only detected.

This attribution problem becomes especially acute in naturalistic settings where behavioral variables are intrinsically correlated. Advances in automated behavior analysis, including unsupervised (MoSeq) \cite{Weinreb2024}, supervised (DeepAction \cite{Harris2023deepaction}, MARS \cite{Segalin2021}), and rule-based (BehaviorDEPOT) \cite{Gabriel2022} approaches, now enable high-throughput extraction of dozens to hundreds of behavioral features from video recordings, producing rich yet highly correlated datasets. In such data a neuron's activity might appear to encode speed when it is actually tuned to the locomotion state, just as a place cell could be misinterpreted as responding to social interaction simply because encounters are tied to specific locations. Correlated variables can thus create apparent mixed selectivity: a neuron tuned to one variable can appear selective for the variables that covary with it. The problem is general in neuroscience: uninstructed movements explain substantial cortex-wide activity \cite{Musall2019}, and theta modulation confounds speed versus phase coding in the hippocampus \cite{Vanderwolf1969, McNaughton1983, Huxter2003}. These examples underscore a general requirement for selectivity analyses: they must explicitly address correlations among behavioral variables and the temporal structure of neural and behavioral signals \cite{NevjenDunn2024}.

Calcium imaging further intensifies these demands by providing large-scale recordings from hundreds to thousands of neurons during continuous behavior \cite{Chen2013, Zong2022}. Unlike spikes, calcium fluorescence is a continuous signal with complex temporal dynamics. Even the fastest indicators (e.g., jGCaMP8) exhibit decay times of tens to hundreds of milliseconds \cite{Zhang2023}. Indicator kinetics cause temporal convolution, where fluorescent signal from earlier events spuriously correlates with later behaviors \cite{Wei2020}. Many spike-based analyses assume point-process observations and do not transfer directly to temporally filtered fluorescence signals, yet deconvolution can introduce model-dependent distortions and may reduce performance in downstream decoding or attribution analyses \cite{Rupprecht2021, Wei2020}. Moreover, behavior is described by variables of fundamentally different types --- from discrete categorical states like grooming or rearing to continuous measures like speed or head direction --- requiring analytical approaches that can handle heterogeneous variable types within a single framework.

Existing approaches typically pursue one of two goals: confirming selectivity to pre-specified variables \cite{o1971hippocampus, mix3, kropff2015speed} or compressing neural-behavior dependencies into latent dimensions \cite{Cunningham2014, Schneider2023cebra}. Neither is built to screen every neuron against every measured variable when behavior is continuous and its variables are correlated. Linear measures on raw variables miss nonlinear dependencies such as place fields, dimensionality reduction obscures the contributions of individual neurons, and probabilistic decoders may predict complex behavior without quantifying which specific behavioral variables explain a given neuron's activity, especially when covariates are correlated \cite{Kriegeskorte2019}.

An information-theoretic approach suits this task because it defines dependence at the level of statistical structure rather than task-specific response models \cite{friston2011functional}, providing a common analytical language for continuous, discrete, circular, and multivariate variables. However, existing toolboxes were not designed for the joint demands of continuous calcium fluorescence and large-scale screening in naturalistic behavior. The most comprehensive, the Neuroscience Information Toolbox (NIT) \cite{Maffulli2022}, provides multiple mutual information (MI) estimators with bias correction and detailed guidance for calcium imaging, but is designed for trial-based experimental paradigms with many repeated trials per condition. Other toolboxes emphasize complementary levels of analysis: MINT for population-level information transmission \cite{MINT2025}; Frites for group-level inference in electrophysiological data \cite{Frites2022}; IDTxl for network connectivity \cite{IDTxl2019}; HOI for higher-order interactions analysis \cite{Neri2024}; and ZETA for parameter-free single-neuron responsiveness testing \cite{Montijn2021zeta}. The Gaussian Copula Mutual Information (GCMI) framework enables efficient MI estimation for continuous data \cite{gcmi}, but MI estimation alone does not resolve key design challenges of free behavior: dealing with autocorrelation and disambiguation of encoding from correlation-driven artifacts. Single-neuron attribution in continuous, free-behavior recordings thus remains out of reach of these tools, statistically and computationally: a valid test of one neuron-behavior pair already requires an MI estimate at thousands of circular shifts and candidate delays.

Here we present INTENSE (INformation-Theoretic Evaluation of Neuronal SElectivity), an open-source tool that redefines selectivity analysis in continuous calcium-imaging data as a discovery problem, enabling efficient systematic identification of diverse neuronal selectivities. INTENSE provides information-theoretic assessment of single-neuron selectivity at scale while accounting for temporal delays, heterogeneous behavioral variables, and correlated behavioral feature spaces. Gaussian-copula estimation of mutual information on representations matched to each variable type, together with permutation testing and delay search accelerated by the fast Fourier transform (FFT), makes this assessment tractable for thousands of neurons, and a disentanglement procedure separates genuine mixed selectivity from selectivity shared by correlated variables. We validated INTENSE on synthetic data with known ground truth and against an established place-cell detection method, and applied it to hippocampal CA1 recordings from freely behaving mice and to an independent dataset from medial entorhinal cortex.

\section{Results}
\label{sec:results}

\subsection{Overview of INTENSE}
\label{sec:overview_intense}

\begin{figure}[htbp]
  \centering
  \includegraphics[width=0.95\linewidth,height=0.62\textheight,keepaspectratio]{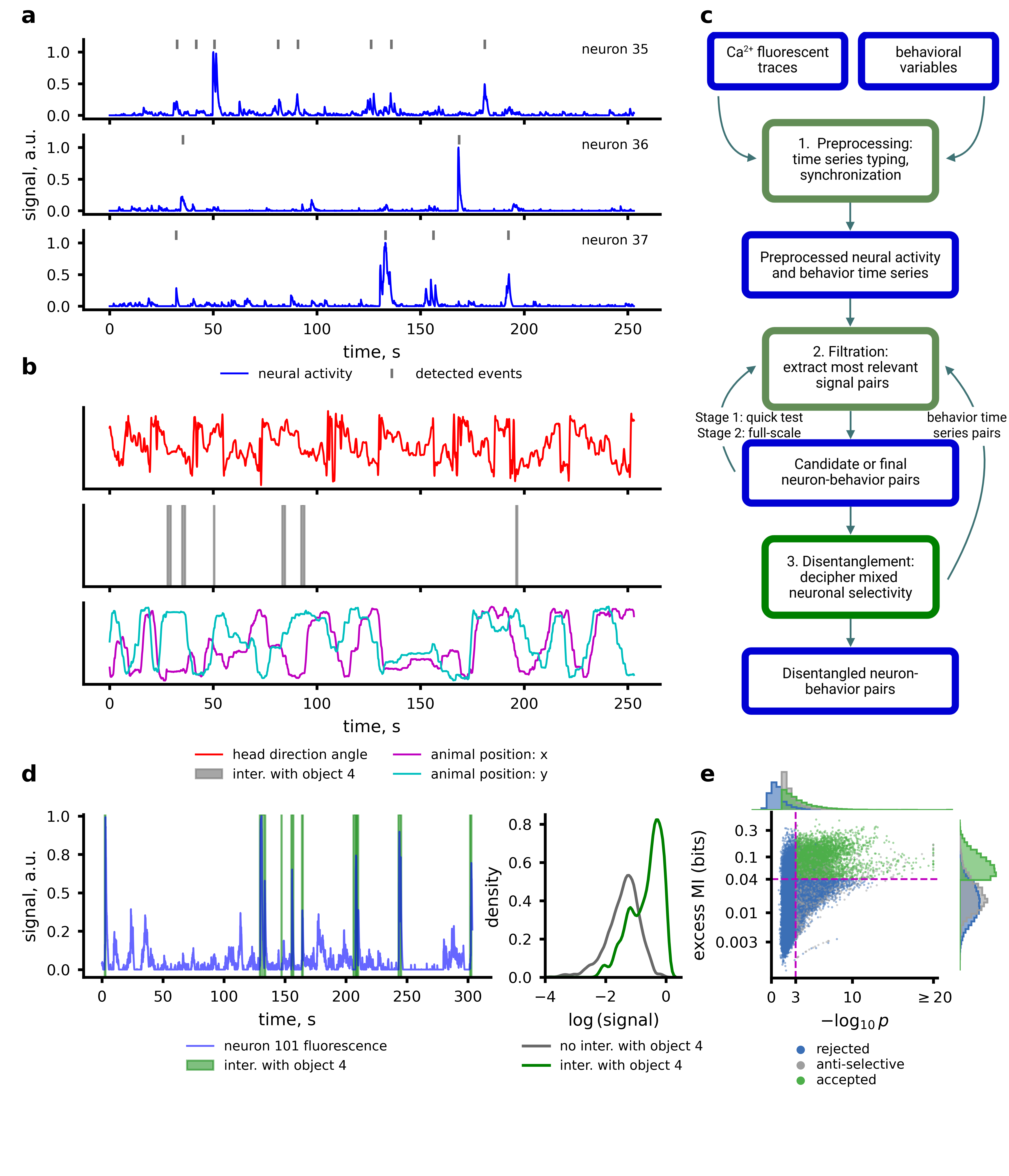}
  \caption{\textbf{Overview of INTENSE.} \textbf{a}, Examples of rescaled calcium fluorescence traces (blue) with detected calcium events (gray ticks).
\textbf{b}, Examples of behavioral/environmental
variables tested against neuronal activity: circular (head direction, top),
discrete (object interaction, middle), multidimensional (animal coordinates,
bottom). \textbf{c}, Schematic overview of INTENSE pipeline for quantifying neuronal selectivity. \textbf{d}, Left -- example object-interaction
cell: fluorescence trace (blue) and interaction periods (green). Right -- distributions of the signal (logarithmic scale) inside (green) and
outside (gray) object interaction periods. \textbf{e}, Distribution of
neuron-behavior pairs of the hippocampal recordings in the ``significance--effect-size''
space with marginal distributions, for pairs that passed the quick first stage of
the test (c): accepted pairs (green), anti-selective pairs, with lower activity during
the behavior, which are not counted as selective (gray), and rejected pairs (blue). Dashed lines
mark the acceptance thresholds ($p = 0.001$ and 0.04 bits; see Methods:
Section~\ref{sec:statistical_assessment}).}

  \label{fig:method}
\end{figure}

The INTENSE toolbox
identifies which neurons carry information about specific behavioral or environmental variables from
large-scale calcium imaging data (\fig{fig:method}). INTENSE works directly with
calcium fluorescence, eliminating the need for spike deconvolution. It uses
mutual information --- which captures any statistical dependency, linear or
nonlinear, equaling zero if and only if variables are independent
\cite{Shannon1948, Borst1999, Strong1998} --- as its core measure. Specifically,
INTENSE employs the Gaussian Copula MI (GCMI) estimator
\cite{gcmi, ma2011mutual}, which provides efficient closed-form MI computation
(see Methods: Section~\ref{sec:mi_computation}), making it feasible to analyze
thousands of neuron-behavior pairs across large populations. Each behavioral
variable enters the estimator in a representation suited to its type, so that
one measure captures the tuning typical of continuous, circular, discrete and
spatial variables, from speed and head-direction tuning to place fields (see
Methods: Section~\ref{sec:mi_computation}).

Statistical significance is assessed
through circular-shift permutation testing with parametric $p$-value modeling
(\fig{fig:method}e; see Methods: Section~\ref{sec:statistical_assessment}), and
systematic delay optimization accounts for calcium indicator kinetics and
prospective or retrospective encoding
(see Methods: Section~\ref{sec:temporal_alignment}). Both steps are accelerated
with the fast Fourier transform, which makes testing every neuron against every
measured variable practical (see Methods: Section~\ref{sec:statistical_assessment}).

Beyond detection, INTENSE provides tools for analyzing mixed selectivity
(\fig{fig:results}f,g). Many hippocampal and cortical neurons show apparent
selectivity to multiple behavioral variables \cite{mix2, mix3, Stefanini2020}, but distinguishing
true mixed selectivity from correlations induced by covarying behaviors is essential
for understanding neural coding principles and the true dimensionality of
representations \cite{mix}. The framework's disentanglement procedure combines priority rules for nested or nearly duplicate
variables, derived from preliminary disentanglement analyses of our behavioral datasets, with
conditional MI and interaction
information \cite{ghassami2017} to separate genuine multi-variable encoding from
spurious associations (see Methods: Section~\ref{sec:disentangling_selectivity};
Section~\ref{sec:intense_deciphers}).

We validated INTENSE on synthetic datasets with known ground truth across a wide
range of signal-to-noise ratios and response reliability conditions
(\fig{fig:synthetics}, Section~\ref{sec:synthetic_validation}), and against
classic event-based place cell detection
(\fig{fig:pc}, Section~\ref{sec:intense_validates}), then applied it to
one-photon calcium imaging data from hippocampal CA1 of freely behaving mice, recovering cells selective
for object interaction and rearing (\fig{fig:results}a,b), position (\fig{fig:pc}b), and both
body direction and position (\fig{fig:results}d,e;
Section~\ref{sec:intense_deciphers}), and to an independent two-photon dataset
from medial entorhinal cortex \cite{Obenhaus2022} (Section~\ref{sec:mec_results}).

\subsection{INTENSE validates against established place cell detection methods}
\label{sec:intense_validates}

\begin{figure}[htbp]
  \centering
  \includegraphics[width=0.9\linewidth,height=0.53\textheight,keepaspectratio]{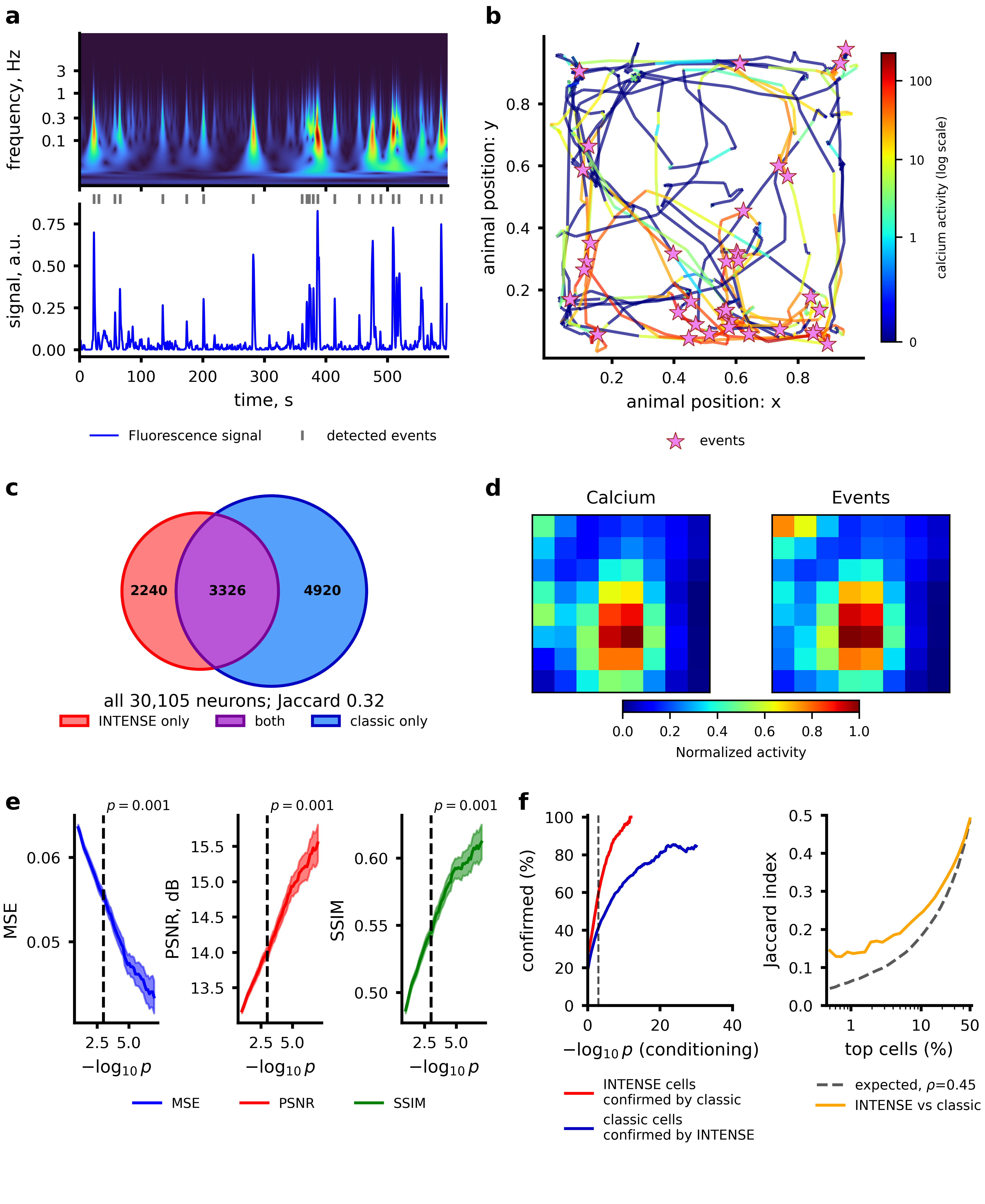}
  \caption{\textbf{Comparison of INTENSE with event-based place-cell detection.} \textbf{a}, Upper -- continuous wavelet transform map (CWT map) of a
fluorescence signal. Cross-scale ridges correspond to detected events. Lower -- fluorescence trace (blue) and detected calcium events (gray ticks). \textbf{b}, Example animal trajectory, colors represent rescaled calcium activity of a single place cell. Stars mark the animal's position at the onset of each detected event. \textbf{c}, Overlap of the
place-cell populations identified by INTENSE (red) and by the classic event-based
method (blue) among all recorded neurons (see Methods: Section~\ref{sec:comparison_pc_populations}). \textbf{d}, Example of activity maps computed from the calcium signal above its 95th percentile (left) and from detected events (right) with a high structural similarity (SSIM) score.
\textbf{e}, Similarity metrics between activity maps computed from the calcium
signal and from detected events as functions of the confidence threshold
applied to both methods (see Methods: Section~\ref{sec:comparison_spatial_selectivity}).
Left -- mean squared error, center -- peak signal-to-noise ratio (dB), right --
structural similarity score. Lines show the mean and shaded regions the s.e.m.
Vertical lines show $p = 0.001$, the per-pair threshold of INTENSE (see Methods:
Section~\ref{sec:statistical_assessment}). \textbf{f}, Agreement between the two
methods as a function of detection confidence (see Methods:
Section~\ref{sec:comparison_pc_populations}). Left -- fraction of cells detected by
one method at a given significance level that are place cells under the other
method (red, INTENSE cells confirmed by the classic method; blue, classic cells
confirmed by INTENSE); dashed vertical line, the same threshold. Right -- overlap (Jaccard index) of the top-ranked cells of
the two methods (orange) as a function of the percentage of cells retained; dashed
line, overlap expected for two statistics with the observed rank correlation.}
  \label{fig:pc}
\end{figure}

To compare INTENSE with a specialized method, we applied it to the well-established task of
identifying place cells (PCs) --- neurons that exhibit selectivity for specific locations in
the environment \cite{o1971hippocampus}.

PC detection provides a challenging validation benchmark, as the choice of
classification method greatly determines the population of cells identified
\cite{Grijseels2021}, with substantial disagreement even between different
event-based methods.

We compared INTENSE, applied directly to calcium fluorescence signals and animal
coordinates (see Methods: Section~\ref{sec:intense_pc_identification}), with a
classic event-based method that tests the spatial information \cite{skaggs1993information}
of event-rate maps built from calcium events detected with a continuous wavelet
transform \cite{Neugornet2021} (see Methods: Section~\ref{sec:classic_pc_identification}).
To compare the two methods directly, we expressed the output of each as (i) a
binary place-cell label and (ii) a confidence score ($p$-value) derived from its
own null model (see Methods: Section~\ref{sec:comparison_pc_populations}).

Over all cells, the two place-cell populations overlapped partly (\fig{fig:pc}c), in
line with the substantial disagreement between place-cell detection methods reported
previously \cite{Grijseels2021}. Their agreement, however, grew with the confidence of
detection: the more significant a cell was for one method, the more likely the other
method also identified it as a place cell, rising from 60\% of the INTENSE place
cells to 94\% of the 100 most significant ones, and from 40\% to 84\% for the classic place
cells (\fig{fig:pc}f).

Significant heterogeneity of PC firing rates and stability has been established both
experimentally \cite{Mizuseki2012, Buzsaki2} and theoretically \cite{Buzsaki2014}.
Our results indicate that INTENSE and event-based methods primarily differ in their
treatment of ``weak'' place cells and agree increasingly on the more reliable ones.
Thus, our general information-theoretic approach successfully captures the spatial
coding properties identified by specialized algorithms.

To verify that place cells identified by both methods share the same pattern of
spatial selectivity, we analyzed the coincidence of activity maps. There exists a
wide variety of place field detection methods \cite{Grijseels2021}, but they are
highly dependent on the data type. To overcome this difficulty, we compared
discretized activity maps (normalized by time spent in the bins; see Methods:
Section~\ref{sec:comparison_spatial_selectivity} and \fig{fig:pc}d).

We quantified the correspondence between activity maps using structural
similarity (SSIM) and complementary metrics (see Methods:
Section~\ref{sec:comparison_spatial_selectivity}).

All spatial correspondence metrics improved with increasing confidence thresholds
(\fig{fig:pc}e), confirming that place cells identified with high confidence by both
methods exhibit highly similar spatial firing patterns. The continuous calcium signal
analyzed by INTENSE therefore carries the spatial tuning revealed by detected events,
so that place cells can be identified and characterized without event detection, a step
that depends on detection parameters and can introduce model-dependent distortions
\cite{Rupprecht2021}.

\subsection{INTENSE reveals diverse neuronal selectivity}
\label{sec:intense_deciphers}

\begin{figure}[htbp]
  \centering
  \includegraphics[width=0.95\linewidth,height=0.66\textheight,keepaspectratio]{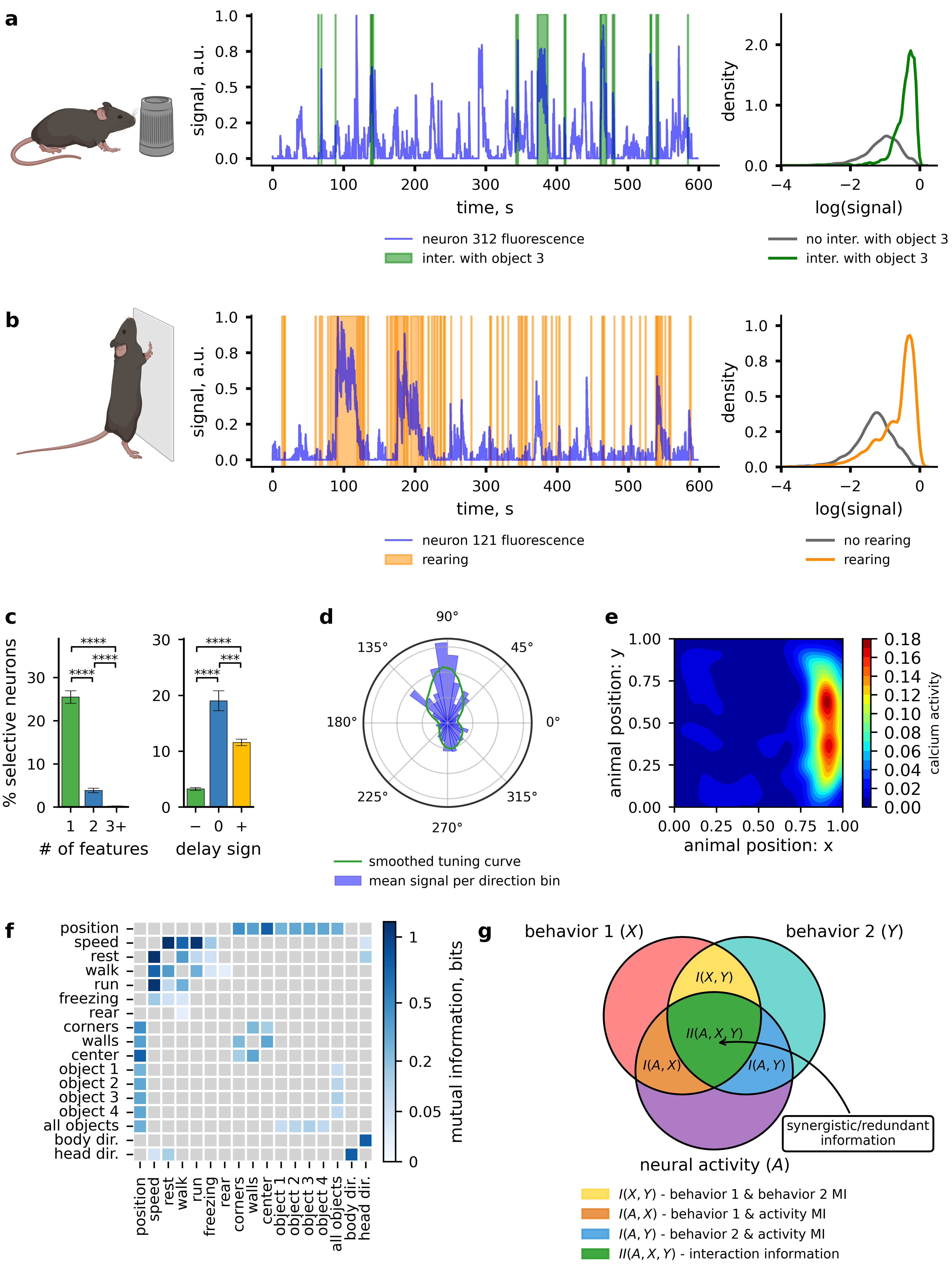}
  \caption{\textbf{Selectivity of hippocampal CA1 neurons and its disentanglement.} \textbf{a}, Example object-interaction neuron: fluorescence (blue)
with periods of interaction with an object (green), and distributions of the signal
(logarithmic scale) during interaction (green) and control periods (gray).
\textbf{b}, Same for an example rearing neuron (rearing periods, orange).
\textbf{c}, Left -- percentage of recorded neurons selective to one, two, or three or
more variables after disentanglement. Right -- selective neuron-feature pairs after disentanglement with
negative ($-$), near-zero and positive ($+$) optimal delays, as a percentage of
recorded neurons. Mean $\pm$ s.e.m. of per-session percentages over $n = 64$ sessions of 16 mice; asterisks, two-sided
Wilcoxon signed-rank tests (see Methods: Section~\ref{sec:statistical_analysis_delays}).
\textbf{d}, Body-direction tuning of a neuron selective for both body direction and
position after disentanglement. \textbf{e}, 2D spatial activity map of the same neuron.
\textbf{f}, Mutual information between behavioral variables, averaged over sessions;
pairs that were not significantly related in most sessions are shown in gray (see
Methods: Section~\ref{sec:disentangling_selectivity}).
\textbf{g}, Schematic of the interdependencies between information measures in
neurons with entangled selectivity. Mouse drawings in \textbf{a},\textbf{b}: Created in BioRender.
Ivashkina, O. (2026) \url{https://BioRender.com/fgprjod}.}

  \label{fig:results}
\end{figure}

Having validated INTENSE against an established method, we applied it to discover
novel neuron-behavior associations in our hippocampal recordings. We analyzed
associations between neural activity and 17 behavioral variables: spatial position,
speed, head and body direction, locomotor states, freezing, rearing, arena zones and object
interactions (see Methods: Sections~\ref{sec:behavioral_analysis} and~\ref{sec:implementation}). Following
identification of significant neuron-behavior pairs, we performed
``disentanglement'' analysis to distinguish true mixed selectivity from spurious
associations arising from behavioral correlations (\fig{fig:results}c,f,g; see
Methods: Section~\ref{sec:disentangling_selectivity}).

We identified neurons selective for various specific behavioral features
(such as object interaction, rearing, etc.; \fig{fig:results}a,b).
These behaviorally selective neurons showed
probabilistic activation patterns --- they neither fired deterministically during
every instance of their preferred behavior nor remained completely silent otherwise (\fig{fig:results}a,b).

Alongside discrete-behavior selectivity, we found numerous cells selective for
continuous variables: movement speed, head and body direction, and spatial position
(\fig{fig:results}d,e). Cells selective for locomotion state, head direction or speed have been increasingly recognized in the
hippocampal-entorhinal circuit \cite{kropff2015speed,Gois2018,Acharya2016}.

\paragraph{Temporal alignment of neural activity and behavior.}
\label{sec:temporal_results}
To account for temporal offsets between behavior and calcium
fluorescence, INTENSE optimizes the delay between signals (see Methods:
Section~\ref{sec:temporal_alignment}).
Optimal delays were binned into three categories (negative, near-zero,
and positive) for statistical analysis. Of the selective neuron-feature pairs, 90.0\%
had near-zero or positive optimal delays, consistent with the expected temporal lag
of calcium indicators, while negative delays, with neural activity preceding
behavior, were the least frequent in every mouse (\fig{fig:results}c; see Methods:
Section~\ref{sec:statistical_analysis_delays}). Because the delay search covers both
signs, INTENSE also detects associations in which neural activity precedes behavior,
as expected for prospective encoding, and these associations are included in all
analyses.

\paragraph{Quantifying and disentangling mixed selectivity.}
Of the recorded neurons (each counted once per session), 25.8\% showed significant
selectivity to at least one behavioral variable, and 19.0\% of these appeared selective for
two or more variables before disentanglement (see Methods:
Section~\ref{sec:statistical_assessment}). Tested against recordings in which the
alignment between neural activity and behavior was destroyed, the same acceptance
criterion was strict enough that only 2--3\% of the accepted pairs are expected to be
false (Extended Data Fig.~\ref{fig:ed_calibration}).

To find the variables that could make a neuron appear selective for more than one of
them, we first measured which behavioral variables were related. Because INTENSE handles
variables of any type, no new procedure was needed. Its two-stage test was run on pairs of
behavioral variables in place of neuron-behavior pairs (see Methods:
Section~\ref{sec:disentangling_selectivity}) and revealed tightly coupled
clusters (\fig{fig:results}f). Locomotion-related variables (speed, rest,
walk, run) were significantly related in 92--100\% of
mouse-session combinations, indicating strong behavioral correlations that could
inflate apparent mixed selectivity. Similarly, spatial zone variables (corners,
walls, center) were significantly related in 89--94\% of mouse-session combinations.

When two behavioral variables are related, a neuron tuned to one of them can also
appear selective for the other. Disentanglement resolves each pair of a neuron's
selectivities by the first of four steps that applies. First, priority rules attribute
the selectivity for labels nested in, or nearly duplicating, another variable (for
example, freezing is part of rest, and locomotor states are defined by speed) to one of
them, so that one selectivity is not counted twice; the rules, derived from preliminary
disentanglement analyses across our behavioral datasets, fix the outcome for these pairs in advance. Two of the rules concern
continuous variables that are strongly coupled during free exploration: locomotor states
take precedence over speed, with which they form a long-recognized confound in the
hippocampus \cite{Vanderwolf1969, McNaughton1983}, and body direction over head direction
(\fig{fig:results}f). Second, if a neuron is selective for both position and a spatial
zone, the two are merged into one place-zone selectivity when its place field lies in
that zone, and otherwise only position is kept. Third, if the two behavioral variables
are not significantly related, both selectivities are kept. Fourth, for related
variables, conditional mutual information (CMI) measures whether the selectivity for one
variable persists once the other is taken into account, and interaction information
(II) tells whether the two carry overlapping (redundant) or complementary (synergistic)
information about the neuron (see Methods: Section~\ref{sec:disentangling_selectivity};
validation of the spatial-zone criterion and the priority rules for objects on synthetic neurons with known selectivity in Supplementary Section~S2).

Disentanglement resolved a large fraction of these multi-variable associations: after
controlling for behavioral correlations, the number of multi-selective neurons
decreased from 1{,}480 to 868 (\fig{fig:results}c), mostly through the priority rules, derived from those preliminary disentanglement analyses,
and the spatial-zone criterion, which build the structure of the behavioral
variables into the procedure (see Methods: Section~\ref{sec:disentangling_selectivity}). Correlations
among behavioral variables therefore explained 41.4\% of the apparent mixed
selectivity.

Interaction information shows how related variables share their information about a
neuron (\fig{fig:results}g). For the 260 pairs of related variables that no rule or zone
criterion resolved, CMI tested whether each selectivity survived conditioning on the other
variable. Both selectivities survived in 92.7\% of these pairs, including most of the
54.2\% of pairs in which the two variables carried partly overlapping (redundant)
information about the neuron: in these pairs each variable carried information about the
neuron that the other did not.

Even after disentanglement, a substantial number of neurons retained significant
associations with multiple variables. For example, the neuron illustrated in
\fig{fig:results}d,e was selective for both body direction and
position. Across the population, more neurons remained multiply selective than expected from the
errors of disentanglement alone. A neuron that codes a single variable can also pass the test for a
variable that covaries with it, and disentanglement does not remove every such label: synthetic
neurons with one true selectivity, placed into the recorded behavior and matched to the selective
CA1 neurons in their variables and information, would keep two or more selectivities in 4.5\% of
cases, compared with 11.2\% of the selective CA1 neurons ($p < 0.001$; Supplementary
Section~S2). While such multiply-selective neurons comprise a minority
of the population (\fig{fig:results}c), they may serve as crucial integration points
linking different functional systems \cite{mix3}.

\paragraph{INTENSE detects associations invisible to linear correlation.}
\label{sec:mi_vs_corr}
To measure what INTENSE gains over linear correlation, we replaced mutual information
with the Pearson correlation between the neural signal and each behavioral variable,
keeping the shuffles and the delay search, and judged both measures by the rank of the
observed value among the shuffles of each pair (see Methods:
Section~\ref{sec:implementation}). Mutual information detected more significant
neuron-feature pairs than correlation, and the two measures detected different pairs:
correlation detected 54\% of the dependencies found by INTENSE, whereas mutual information
detected 82\% of the same number of the strongest correlation pairs. The difference was largest for
head and body direction, where correlation detected 30\% of the dependencies found by
INTENSE. A substantial fraction of
neuron-behavior dependencies is thus invisible to linear correlation with the raw variables,
and INTENSE recovers it by measuring information on representations suited to the type of
each variable, consistent with the synthetic recordings with known dependencies, where INTENSE
reached a higher $F_1$ score than correlation with shuffles on continuous variables
(Section~\ref{sec:intense_outperforms}).

\paragraph{Application to an independent dataset.}
\label{sec:mec_results}
To test INTENSE beyond our own recordings, we applied it to an independent
two-photon dataset from the medial entorhinal cortex of freely moving mice, recorded
in another laboratory \cite{Obenhaus2022}, in which the dataset authors had labeled
functional cell types. We compared INTENSE with the generalized linear model (GLM),
a standard approach to attributing a neuron's activity to several behavioral variables
\cite{hardcastle2017multiplexed}. INTENSE recovered more than twice as many of the
head-direction cells labeled by the authors as a forward-selection GLM (77\% versus
29\%, without the 0.04-bit threshold; Supplementary Section~S1.5.1). INTENSE thus transfers to another brain region,
imaging method and laboratory.

\subsection{Synthetic validation demonstrates INTENSE robustness}
\label{sec:synthetic_validation}

\begin{figure}[htbp]
  \centering
  \includegraphics[width=\linewidth,height=0.6\textheight,keepaspectratio]{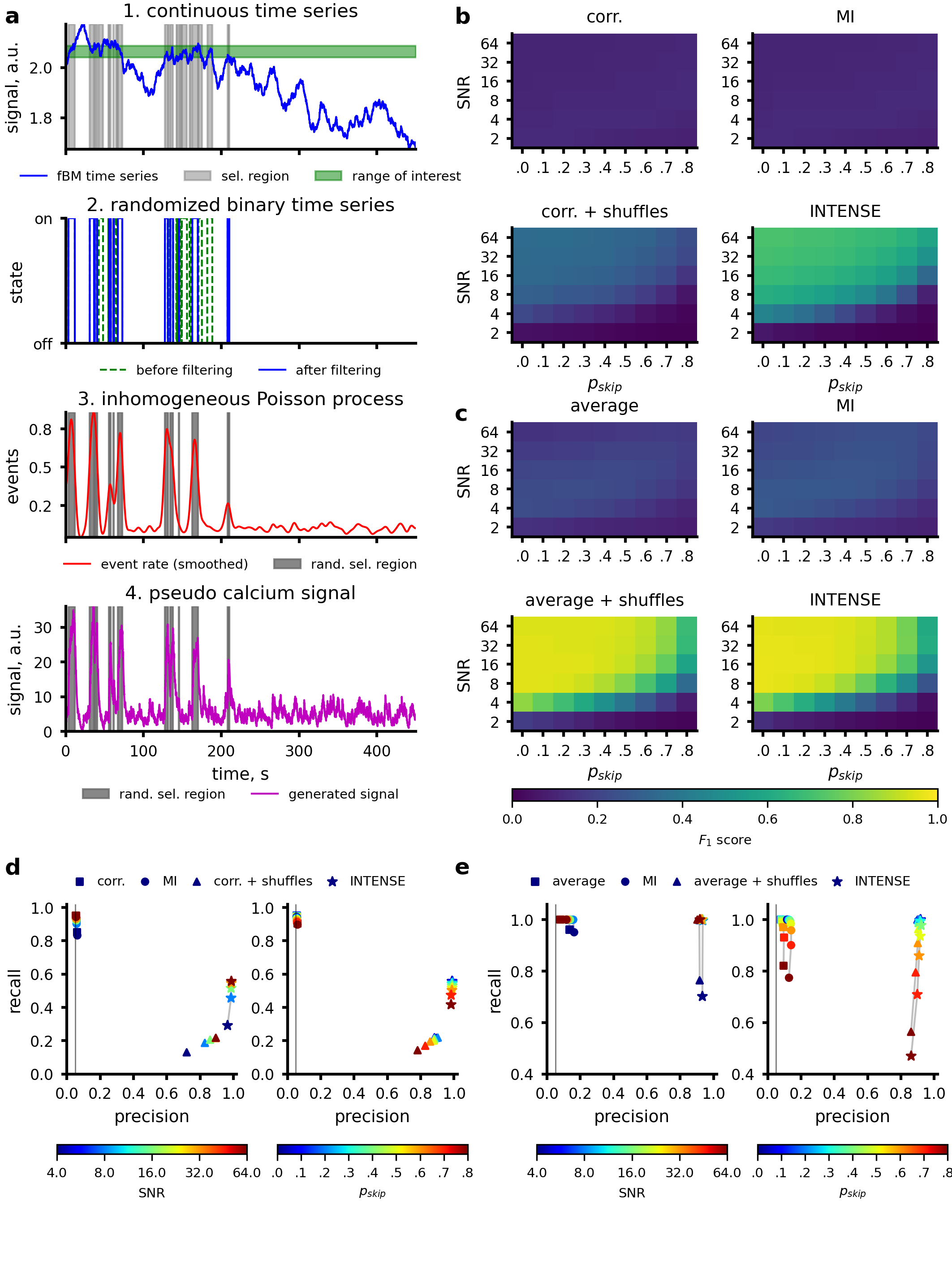}
  \caption{\textbf{Detection of neuron-behavior dependencies in synthetic data with known ground truth.}
\textbf{a}, Stages of generating a pseudo-calcium signal associated with a continuous
variable: 1 -- continuous variable (a fractional Brownian motion, fBM), the neuron's
range of interest (green) and the periods the variable spends in it (gray); 2 -- these
periods as a binary time series, before and after filtering, which deletes whole
periods with probability $p_{\text{skip}}$; 3 -- event rate of the inhomogeneous
Poisson process (smoothed); 4 -- resulting pseudo-calcium signal. For discrete variables, the procedure starts
from stage~2.
\textbf{b}, $F_1$ score over the grid of SNR and $p_{\text{skip}}$ for continuous
variables: correlation and MI with analytical tests, correlation with shuffles, and
INTENSE. \textbf{c}, Same as b for discrete variables: difference of mean activity
(average) and MI with analytical tests, ratio of mean activity with shuffles
(average + shuffles), and INTENSE.
\textbf{d}, Precision and recall of all methods for continuous variables at different
SNR values ($p_{\text{skip}} = 0$, left) and different $p_{\text{skip}}$ values
(SNR $= 64$, right); vertical lines, precision of a random guess (0.05).
\textbf{e}, Same as d for discrete variables. Means over 5 realizations of each
condition (see Methods: Section~\ref{sec:experimental_protocol}).}
  \label{fig:synthetics}
\end{figure}

\subsubsection{Synthetic framework and the role of temporal controls}
\label{sec:bio_synthetic}

To rigorously evaluate INTENSE's detection capabilities, we conducted systematic
tests on realistic synthetic datasets with known ground truth, in which calcium signals were
linked to behavioral variables through a biologically plausible process
(\fig{fig:synthetics}a; see Methods: Sections~\ref{sec:synthetic_data_generation} and~\ref{sec:signal_generation}).
Each experiment contained 20 behavioral time series and 500 pseudo-calcium signals,
with each calcium signal linked to exactly one behavioral variable (5\% prevalence of
true associations), and the framework varied the signal-to-noise ratio (SNR, the ratio of firing rates inside versus
outside the neuron's target regions, from 2 to 64) and response reliability (probability $p_{\text{skip}}$ of skipping a response,
from 0 to 0.8). INTENSE was compared with analytical tests of correlation (continuous
variables) or of mean activity (discrete variables) and of MI, and with correlation or
activity ratio tested within the INTENSE procedure, with the same shuffles and delay
search; all methods used the same per-pair threshold, $p < 0.001$ (see Methods:
Section~\ref{sec:comparison_methods}).

A method comparison revealed a fundamental challenge: temporal autocorrelation
creates overwhelming false positive rates without proper statistical controls
(\fig{fig:synthetics}b--e). Analytical tests, which treat the samples of calcium
signals as independent, detected most true pairs, but, on average across conditions, only 6--13\% of their
detections were true, barely exceeding the 5\% expected by chance.
This demonstrates why shuffle-based significance testing is essential --- the slow
dynamics of both calcium signals and behavioral variables create spurious
correlations that appear highly significant under naive statistical tests.

\subsubsection{INTENSE outperforms correlation-based shuffle approaches}
\label{sec:intense_outperforms}

Among methods employing shuffle controls, INTENSE demonstrated clear advantages over
correlation on continuous variables: it achieved the highest $F_1$ score in 47 of the
54 conditions and, with reliable responses and strong signals, detected more than twice
as many true pairs as correlation with shuffles, at a precision of 0.99
(\fig{fig:synthetics}b,d). Synthetic neurons responded within a
limited range of their variable, as place and head-direction cells do, so most
dependencies were non-monotonic; INTENSE captures such tuning through the
representation of continuous variables, whereas correlation registers only its linear
component (see Methods:
Section~\ref{sec:mi_computation}). This pattern reveals a fundamental limitation:
correlation-based methods, even with proper statistical controls, remain restricted
to linear associations and miss the complex nonlinear dependencies often present in
neural data.

For discrete variables, where the dependence is a difference in activity between two
states, INTENSE matched the activity-ratio test with shuffles, and with reliable
responses both detected nearly all true pairs at moderate and high signal-to-noise
ratios (\fig{fig:synthetics}c,e). The general measure of INTENSE thus performs on a
par with a statistic tailored to this simpler form of dependence.

\subsubsection{Two-stage testing balances efficiency and power}
\label{sec:two_stage_testing}

Because the full test is costly, every pair is first screened with a quick test on a
small number of shuffles, and only the candidates that pass it receive the full test
(see Methods: Section~\ref{sec:statistical_assessment}). This two-stage procedure
proved highly effective at managing the computational burden of large-scale
analysis. Stage~1 retained 2--8\% of the pairs as candidates, and Stage~2 removed most
false candidates: with the strongest signals and reliable responses, precision rose
from 0.58 and 0.61 among the candidates to 0.99 and 0.92 among the accepted pairs,
for continuous and discrete variables respectively (see Methods:
Section~\ref{sec:experimental_protocol}), demonstrating effective false discovery
control while the full test was computed for only a small fraction of the pairs.

\subsubsection{Performance under biological variability}
\label{sec:performance_variability}

Real neurons show variable response reliability --- they may skip responses due to
adaptation, internal state changes, or stochastic factors. INTENSE remained precise as
responses became less reliable: with strong signals, its precision on continuous
variables stayed at 0.98--0.99 at every skip probability while its recall declined
gradually, and on discrete variables it kept both precision and recall high ($F_1$ score of 0.95 or above) up to a skip
probability of 0.3
(\fig{fig:synthetics}d,e), demonstrating robustness to moderate response unreliability
commonly observed in real neural data.

A notable pattern emerged at high signal-to-noise ratios: on discrete variables, the
precision of analytical tests decreased as the signal grew stronger, whereas
shuffle-based tests, including INTENSE, were nearly unaffected (\fig{fig:synthetics}c,e). We
attribute this to the overlap of slow calcium responses: when neurons fire
frequently, slow decay dynamics keep fluorescence elevated between events, merging
individual episodes into sustained plateaus that tests assuming independent samples
read as dependence. Circular-shift shuffles construct the null distribution from the
same signals, so this overlap elevates both true and null MI values alike,
preserving the validity of the significance test.

These synthetic results demonstrate that INTENSE successfully addresses the key
challenges in detecting neuron-behavior associations: controlling false discoveries
from temporal structure, capturing complex associations beyond linear correlation,
and maintaining sensitivity despite biological variability. The framework's robust
performance across diverse conditions establishes confidence for its application to
experimental data where ground truth is unknown.

\section*{Discussion}
\label{sec:discussion}

We recast selectivity analysis in calcium imaging as a discovery problem: INTENSE screens every neuron against every measured behavioral variable on one information-theoretic scale, working directly on fluorescence and testing significance against nulls that preserve its temporal structure. Specialized pipelines for place, head-direction or speed cells each handle one variable type. A common measure makes selectivities of different types comparable within the same neuron, which is what quantifying mixed selectivity and separating it from behavioral covariance requires. Mutual information computed on representations matched to each variable type covers continuous variables such as speed, circular ones such as head direction, discrete states such as rearing or object interaction, and two-dimensional position.

INTENSE combines GCMI with circular-shift permutation testing to detect neuron-behavior dependencies in calcium imaging data, achieving balanced performance across a wide range of signal conditions (\fig{fig:synthetics}b,d,e). When applied to the classic benchmark of place cell identification, INTENSE agreed with an established event-based method \cite{skaggs1993information} on the most reliable place cells, and the agreement grew with the confidence of detection (\fig{fig:pc}). The framework operates directly on calcium fluorescence without requiring spike deconvolution, avoiding potential distortions or information loss introduced by preprocessing \cite{Rupprecht2021, Wei2020}. The rank-based GCMI estimator manages baseline variations and non-Gaussian transient shapes typical of calcium signals \cite{gcmi}, while two-stage testing enables large-scale analysis by screening out non-significant pairs early. Importantly, synthetic benchmarks underscore a critical requirement for calcium imaging analysis: significance testing requires accounting for temporal autocorrelation (\fig{fig:synthetics}). Without appropriate shuffle controls, the slow dynamics of calcium indicators can inflate apparent selectivity and produce false discoveries \cite{NevjenDunn2024}. Moreover, circular-shift null distributions remain valid at high signal-to-noise ratios, where tests without temporal controls make more false detections as the signal grows stronger (\fig{fig:synthetics}e).

Applied to hippocampal activity data during free open field exploration, INTENSE recovers results consistent with key principles of CA1 coding while highlighting the limits of commonly used linear measures. Consistent with sparse coding, only a minority of neurons show significant selectivity to any given behavioral variable, in line with estimates of environment-specific recruitment in CA1 \cite{ThompsonBest1989, Epsztein2011}. Moreover, linear correlation in the same procedure missed about half of the dependencies INTENSE detected (Section~\ref{sec:mi_vs_corr}), confirming that a substantial fraction of neuron-behavior dependencies are invisible to linear measures. This observation aligns with recent work emphasizing that neural representations occupy curved, nonlinear subspaces, motivating analytical tools that do not assume simple parametric coupling \cite{De2023, intdim}. On an independent dataset from medial entorhinal cortex, recorded with another imaging method in another laboratory, INTENSE recovered the head-direction cells labeled by the dataset authors more completely than a GLM did (Section~\ref{sec:mec_results}).

A major goal of INTENSE is to shift selectivity analysis from confirmation of preselected cell classes to principled discovery and adjudication of neuron-behavior dependencies in correlated behavioral state spaces. Temporal autocorrelation can make naive tests report associations that are absent, and correlations among behavioral variables can generate associations that are statistically real yet misleading with respect to what a neuron encodes \cite{Harris2021nonsense, NevjenDunn2024}. More broadly, behavioral variability and uninstructed movements can dominate neural activity and confound selectivity estimates \cite{stringer2019spontaneous, Musall2019}. INTENSE addresses these concerns through multivariate information analysis that separates true mixed selectivity from apparent selectivity arising from behavioral correlations. Conditional mutual information tests whether selectivity to a variable persists after controlling for a correlated behavior \cite{gcmi}. Interaction information further provides a compact summary of whether variables contribute redundantly or synergistically, helping to distinguish apparent mixed selectivity driven by covariance from encoding that reflects conjunction-like computation. This disentanglement becomes particularly important for calcium imaging, where lower temporal resolution already reduces apparent mixed selectivity compared to electrophysiology \cite{Wei2020}. An alternative approach uses encoding models such as generalized linear models (GLMs) to simultaneously regress neural activity against multiple covariates \cite{hardcastle2017multiplexed}. GLMs offer simultaneous covariate control but assume a parametric link function and require type-specific encoding models, whereas MI-based conditioning is nonparametric, applies uniformly across continuous, circular, and discrete variables, and captures nonlinear dependencies at the cost of pairwise rather than simultaneous control. In a direct comparison of detection on synthetic recordings with independent behavioral variables, INTENSE detected linear mixed selectivity with an $F_1$ score more than twice that of the strongest GLM variant, without false detections by either method (Supplementary Section~S1.2).

Applying this disentanglement procedure, INTENSE reliably identifies established selective neurons --- place cells, head direction cells, object cells --- while evaluating all of them on a common statistical footing. At the same time, the framework refines estimates of mixed selectivity in naturalistic conditions: about one in five selective neurons initially appeared multi-selective, but disentanglement attributed about 40\% of these to correlations among behavioral variables (Section~\ref{sec:intense_deciphers}); the mixed selectivity that remains is consistent with distributed mixed coding in the hippocampus \cite{McNaughton1983, Rubin2015, Sarel2017}. Some variables remain fundamentally ambiguous under common experimental constraints: when objects occupy fixed locations, place and object selectivities cannot be clearly separated \cite{Plusnin2021}, and such cases may plausibly reflect genuine conjunctive encoding. More broadly, these results suggest that estimates of mixed selectivity in hippocampus obtained without explicit control for behavioral covariance may conflate genuine multivariable dependence with structure already present in behavior \cite{TyeMiller2024}.

These findings are compatible with theoretical perspectives proposing that apparent complexity can arise from a combination of sparse selectivity at the single-neuron level and structured population codes \cite{TyeMiller2024}. Although neurons often appear mixed-selective under traditional analysis, our pairwise analysis indicates that most carry information about a single variable after controlling for behavioral covariance. The implications of this single-neuron sparsity for population-level coding geometry remain to be tested directly \cite{Stefanini2020}. The observed sparsity aligns with the log-dynamic brain hypothesis \cite{Buzsaki2014}, where networks organize around minorities of highly active neurons \cite{Buzsaki2}, whose mixed selectivity supports flexible integration across variables \cite{mix3}. Analogous motifs are discussed in deep artificial networks, where mixed-selective units form low-dimensional representations that support flexible, abstract computation \cite{Bernardi2020, Johnston2023}. While the biological and artificial settings are not directly comparable, these parallels motivate the broader hypothesis that sparse mixed selectivity may be a computational strategy that balances efficiency with representational power.

Although two-stage testing and FFT acceleration improve scalability, information-theoretic analyses remain computationally demanding as the number of neurons, behavioral variables, and candidate interactions grows. In addition, our disentanglement procedure can only control for measured covariates; unobserved or poorly captured behavioral factors can still induce residual confounds. The probabilistic activation patterns observed in selective neurons suggest that behavioral acts appearing identical from kinematic observation may carry distinct contextual significance for the animal, representing one such unmeasured source of variability. The disentanglement procedure also decides whether two behavioral variables are related by comparing their coupling at the same moment with that after one of them is circularly shifted. A coupling that unfolds over seconds, such as between head or body direction and the position the animal reaches shortly afterwards, can therefore be missed, and part of the selectivity for both position and direction may reflect such coupling. Extending disentanglement to such couplings is left for future work. The choice of the GCMI estimator represents a deliberate trade-off between nonlinearity detection and computational tractability. With the type-based representations, GCMI captures the tuning typical of each variable type, such as single-peaked tuning to continuous variables, tuning to circular variables and place fields, including those of cells with several fields, and enables FFT-accelerated permutation testing and copula reuse across shifts, making large-scale screening feasible. Non-parametric estimators such as the k-nearest neighbor (KSG) approach \cite{Kraskov2004}, though supported in INTENSE, can capture more complex dependencies but require sample sizes that are prohibitive in shuffle-heavy pipelines and become even more expensive when local non-uniformity correction is applied \cite{Gao2015LNC}. Finally, calcium indicator kinetics and imperfect temporal alignment between neural signals and behavioral variables impose practical constraints on lag selection and temporal resolution, which can influence estimates of selectivity and redundancy. Future integration of Partial Information Decomposition could provide a more granular partition of information into unique, redundant, and synergistic components across multiple variables, strengthening mechanistic interpretation of mixed selectivity beyond pairwise and conditional tests \cite{Williams2010}. INTENSE also makes explicit the search space of selectivity, which spans many behavioral dimensions \cite{stringer2019spontaneous}: its output is a structured map of statistically supported neuron-variable relations that can be interpreted, grouped and tested experimentally.

Overall, INTENSE provides a discovery-oriented information-theoretic toolbox for continuous calcium-imaging datasets, designed to reveal neuron-behavior dependencies without restricting analysis to predeclared cell types and to distinguish genuine multivariable selectivity from covariance-driven artifacts. Its generality across neural signal and behavioral variable classes makes it well suited for integration with modern automated phenotyping pipelines. By disentangling behavioral confounds, the method indicates that computational complexity is likely to arise from small, sparse populations of mixed-selective neurons, rather than from widespread conjunctive encoding across the neural population. As recordings expand to larger populations and increasingly rich behavioral paradigms, correlation-aware selectivity analyses will be essential for making interpretable claims about neuron-variable relations and for avoiding conflation of genuine selectivity with correlation artifacts.

\bibliographystyle{elsarticle-num}
\bibliography{main}

\section{Methods}
\label{sec:methods}

\subsection{Experimental setup}
\label{sec:experimental_setup}
We identified behaviorally selective neurons in mice during free exploration of a square open field arena enriched with a variety of contextual cues. Mice were repeatedly placed into the environment for exploration while their neural activity was recorded using miniscope in vivo calcium imaging.

\subsubsection{Animals}
\label{sec:animals}
We used C57BL/6J mice (n=16; 9 males and 7 females; bred in-house from Jackson Laboratory stock, strain \#000664), aged 6.0--8.5 months (184--258 days) on the recording days. Before surgery, animals were housed in standard laboratory cages (2--7 per cage) under a 12-hour light/dark cycle, at 22--23~\textdegree{}C and 45--50\% humidity, with free access to food and water. After surgery, mice were housed individually; other housing conditions remained unchanged. All experiments were performed during the light phase of the day (10:00 a.m. to 6:00 p.m.).

\subsubsection{Surgeries and viral delivery}
\label{sec:surgeries}
All procedures were approved by the Commission of Bioethics of Lomonosov Moscow State University (Application \textnumero{} 159-g, approved during Bioethics Commission meeting \textnumero{} 154-d-r held on 17.08.2023), and followed the Russian Federation Order N 267 MZ and the NIH Guide for the Care and Use of Laboratory Animals.
Surgical procedures were carried out in three stages: viral injection, GRIN lens implantation, and baseplate installation. Anesthesia was induced by intraperitoneal injection of zoletil (40 mg/kg) and xylazine (5 mg/kg). Dexamethasone (4 mg/kg) was administered subcutaneously 5 minutes before surgery to reduce inflammation and prevent cerebral edema. To prevent eye drying, we applied moisturizing gel. Mice were positioned in a stereotaxic frame (Kopf, USA), and body temperature was maintained with a heated pad (Physitemp, USA). For virus injection, 300 nL of AAV/DJ-CAG-GCaMP6s (Federal Medical-Biological Agency of Russia) was delivered unilaterally into the dorsal CA1 region (coordinates: AP -1.94 mm, ML +1.46 mm, DV -1.2 mm, \cite{Paxinos2019}). One week later, a 1-mm-diameter GRIN lens (GrinTech, Germany) was implanted 200 $\mu$m above CA1. The lens was attached to the skull using cyanoacrylate glue (Henkel, Germany) and dental cement (Stoelting, USA), and covered with Kwik-Sil (World Precision Instruments, USA). In the third surgery, performed one week later, an aluminum baseplate was placed over the implant site to allow miniscope mounting and fixed in place using dental cement. A plastic cap was added to protect the lens. After the final surgery, mice were returned to their home cages and given three weeks to recover. Lens placement in CA1 was verified histologically in all mice.

\subsubsection{Behavioral procedure}
\label{sec:behavioral_procedure}
After recovery, mice were gradually habituated to handling and wearing a miniscope. For three consecutive days, each mouse was placed in its home cage with a connected but non-recording miniscope for 5 minutes per session. This step helped animals adapt to the weight and presence of the device without additional stress. Following habituation, animals were tested in a square open field arena ($44 \times 44 \times 44$~cm, gray plastic; Open Science, Russia) for four consecutive days (64 sessions from 16 mice). Calcium imaging was recorded at 19.8 frames per second in 26 sessions and at 29.8 frames per second in 38 sessions, and all 64 synchronized recordings were about 10 minutes long. The arena was designed to provide a rich contextual environment, including a semi-transparent bubble silicone floor mat and a distinct arrangement of four floor objects: a white Styrofoam cone, a cell culture flask filled with bedding, a blue wooden cylinder, and a yellow plastic construction block shaped like a parallelepiped with a hole. Each wall of the arena featured a unique visual cue: (1) a yellow circle with five small blue dots, (2) a blue triangle, (3) a black square with wide white stripes, and (4) a composite figure consisting of blue trapezoids with a black inverted triangle and a white circle in the center. Throughout each trial, animal behavior was recorded using a top-mounted video camera (Flir Chameleon3, KVR, UK), while CA1 calcium activity was simultaneously acquired with a Miniscope V4.4 (Open Ephys). Behavioral video and miniscope streams were synchronized in Bonsai version 2.7.2 (Bonsai Foundation, UK) using shared timestamps, yielding aligned time bases for delay-dependent analyses. Neural and behavioral time series of each session were aligned to the frames of its calcium recording with the IABS data synchronizer (version 1.1.0;
\href{https://github.com/iabs-neuro/iabs_synchronizer}{https://github.com/iabs-neuro/iabs\_synchronizer}), which interpolates
each behavioral variable linearly onto the calcium timestamps.

\subsubsection{Behavioral analysis}
\label{sec:behavioral_analysis}
Behavioral tracking and annotation were performed using the Sphynx package for exploratory behavior analysis \cite{Plusnin1}. For each video, the positions of key body parts were estimated frame-by-frame using a neural network trained with DeepLabCut version 2.3 \cite{Mathis}. Coordinates with confidence scores below 0.95 were excluded and reconstructed by piecewise cubic interpolation using neighboring high-confidence frames. The resulting trajectories were smoothed with a third-degree Savitzky--Golay filter, applied with a 0.25-second window for the body center and tail base, and 0.1 seconds for all other body parts. The arena was segmented into functionally distinct spatial zones \cite{Plusnin1}: $8.5 \times 8.5$~cm corner zones, 6-cm-wide wall zones along the perimeter between the corner zones, a center zone covering the rest of the arena floor, and 2.5-cm object zones surrounding each object. Both continuous and discrete behavioral variables were calculated based on the animal's movement and position relative to these areas. Continuous variables included:
Cartesian coordinates of the body center;
speed (absolute velocity), calculated from body center displacement and smoothed with a 0.25-second window;
body direction (BD), defined as the angle from tail base to body center;
head direction (HD), defined as the angle from head center to the miniscope, smoothed with a one-second Savitzky--Golay filter.
Discrete behavioral acts were classified based on movement dynamics and spatial transitions. Locomotor states were defined by current speed: run (>5 cm/s), walk (1--5 cm/s) and rest (<1 cm/s). Freezing was defined as simultaneous decrease of both body center velocity (<1 cm/s) and nose tip velocity
(<2 cm/s). Rearing was detected by a transient reduction in the distance between the
hind limbs and the tail base. Presence in the corner, wall and center zones was determined
by the position of the tail base, and object interactions by the position of the nose tip. All discrete
variables were smoothed using a 0.25-second median filter to remove noise and
enforce minimal event duration.

\subsubsection{Processing of calcium imaging data}
\label{sec:calcium_processing}
Calcium imaging data were processed using BEARMIND (\href{https://github.com/iabs-neuro/bearmind}{https://github.com/iabs-neuro/bearmind}), a custom analysis pipeline
developed by the authors, built on the CaImAn framework (version 1.11.3)
\cite{Pnevmatikakis2016, Giovannucci2019}. The software was designed to support end-to-end processing of multi-session miniscope recordings.
Each session began with visual inspection to define cropping boundaries.
Preprocessing steps included background correction and motion artifact removal using the NoRMCorre algorithm \cite{Pnevmatikakis2017}. Videos were then decomposed into spatial footprints and temporal activity traces using constrained non-negative matrix factorization (CNMF-E) \cite{Zhou2018}. CNMF-E hyperparameters were selected using predefined quality criteria (spatial footprint compactness, trace SNR, and residual structure) and verified by visual inspection. The main parameter ranges were:
\begin{itemize}
\item \texttt{gSig} (estimated neuron size): 3--6 pixels
(with resolution 1.5 $\mu$m/pixel),
\item \texttt{min-corr} (local correlation threshold):
0.85--0.95,
\item \texttt{min-pnr} (peak-to-noise ratio): 3--15.
\end{itemize}
After decomposition, putative components were automatically filtered using
thresholds \texttt{rval-thr} = 0.95 and \texttt{min-SNR} = 3, following the
\href{https://caiman.readthedocs.io/en/latest/CaImAn\_Tips.html}{CaImAn guidelines}.
All retained components were visually inspected using BEARMIND to identify and
exclude artifacts and duplicates. Only components verified by this manual step were
classified as neurons.

The denoised calcium traces of the retained components, estimated by CNMF-E with a
first-order autoregressive model of calcium dynamics, were used for analysis.

\subsubsection{Calcium event detection}
\label{sec:calcium_event_detection}
To extract discrete calcium transients from the calcium fluorescence traces, we used a
wavelet-based approach previously described in \cite{Neugornet2021}, with minor
modifications. This method was chosen for its robustness to noise and its ability to
detect multiscale features in fluorescence data.

The fluorescence trace was first smoothed by convolution with a Gaussian kernel ($\sigma =
8$ frames), then transformed using a continuous wavelet transform (CWT), yielding a
two-dimensional array of wavelet coefficients across time and scale. Significant
events appeared as ``ridges'' of local maxima across adjacent scales (\fig{fig:pc}a).

We used generalized Morse wavelets $\Psi_{\beta,\gamma}(\omega) =
\alpha_{\beta,\gamma}\omega^{\beta}e^{-\omega^{\gamma}}$ with parameters $\gamma =
3$ and $\beta = 2$, corresponding to so-called Airy wavelets, which offer a minimal
Heisenberg area and good temporal localization.
For each scale, local maxima of the wavelet coefficients (real part) were detected.
Maxima at the smallest scale initiated ridges, which were then extended to larger
scales when a maximum of the next scale fell within the time resolution of the wavelet.
When multiple candidates were present at a given step, only the highest local maximum
extended the ridge, while others initiated new ridges. Ridges present in fewer than 40
of the 50 scales were discarded.
Once all ridges were constructed, a significance filter was applied based on ridge
duration, peak amplitude, and the scale at which the maximum occurred. Ridges that
passed this filtering step were classified as calcium events. Each event was represented by the interval spanned by its ridge,
limited to 0.40--2.5~s (0.47--2.5~s in the sessions recorded at about 30 frames per
second); overlapping intervals were merged, and the first frame of each interval was taken
as the event onset. Events were detected with the implementation in DRIADA 1.2.2
\cite{driada}.

\subsection{INTENSE pipeline}
\label{sec:intense_pipeline}

\subsubsection{Overview of the procedure}
\label{sec:implementation}

The hippocampal recordings were analyzed with INTENSE as implemented in DRIADA version
1.2.2 \cite{driada}. In each recording session, every neuron was tested against 17
behavioral features: spatial
position, speed, body direction and head direction, and the discrete variables rest,
walk, run, freezing, rearing, walls, corners, center, objects 1--4 and pooled
objects; the coordinates $x$ and $y$ entered only through spatial position. The analysis
of each session had three steps. First, each behavioral variable was represented
according to its type, and its dependence on the calcium signal was measured by mutual
information (Section~\ref{sec:mi_computation}). Second, the significance of each
neuron-feature pair was assessed against circular shifts of the behavior, at the delay
that maximized mutual information (Sections~\ref{sec:statistical_assessment}
and~\ref{sec:temporal_alignment}).
A pair was accepted as selective if its $p$-value was below 0.001, without correction
for multiple comparisons, its information exceeded the chance level by more than
0.04~bits, and neural activity was higher in the preferred range of the variable
(anti-selectivity filter, described with the exceptions to it in
Section~\ref{sec:statistical_assessment}); optimal delays of all signs were kept. Third, for every
neuron with at least two accepted features, disentanglement decided which of them were
retained and which were attributed to correlations among behavioral variables
(Section~\ref{sec:disentangling_selectivity}). Only accepted pairs took part in it, so
that a pair below the information threshold could not affect a decision about the
neuron's selectivities. The accepted pairs of all sessions were then collected for the counts and
figures; a merged place-zone selectivity was counted under its zone label.

\textbf{Comparison with linear correlation.} For the comparison of MI with linear
correlation (Section~\ref{sec:mi_vs_corr}), the same analysis, with the same shuffles and
delay search, was run with the absolute value of the Pearson correlation, $|r|$, as the
dependence measure, computed between the neural signal and each behavioral variable in its
original form. Spatial position was not included in this comparison. Both measures were
judged by the rank of the observed value among the Stage~2 shuffles
(Section~\ref{sec:statistical_assessment}), at $p < 0.001$, with the anti-selectivity filter
and before disentanglement. The rank was used for both measures because the zero-inflated
gamma model that gives the parametric $p$-value of MI overstates the tail of the shuffled
values of $|r|$ more than that of MI: at the value exceeded by 1\% of the shuffles, the
fitted model gave a median tail probability, over pairs, of 1.1\% for MI and 2.7\% for
$|r|$, so the parametric test would have been stricter for correlation. By the rank, before
the excess-information threshold, MI detected 22{,}402 significant pairs and correlation
19{,}040. To compare which pairs the two measures detect, we took the dependencies found by
INTENSE (the 5{,}297 significant MI pairs whose information exceeded the chance level by more than
0.04~bits) and the same number of the strongest significant correlation pairs. Correlation
pairs were ranked by their effect on the scale of MI for a Gaussian pair,
$-\tfrac{1}{2}\log_2(1 - r^2)$, minus the same quantity for the mean of the shuffled values.
A pair was counted as detected by the other measure if it was significant for that measure
by the rank; the 0.04-bit threshold only defined the dependencies found by INTENSE. When
correlation was measured instead on the type-based representations of INTENSE, between the
neural signal and its best linear prediction from the components of the representation (the
multiple correlation coefficient $R$), it detected 75\% of the dependencies found by
INTENSE, and MI detected 80\% of the strongest correlation pairs.

\subsubsection{MI computation}
\label{sec:mi_computation}

We used the Gaussian Copula Mutual Information (GCMI) method
\cite{gcmi, ma2011mutual} to estimate MI. This method takes advantage of the
property that mutual information depends only on the dependency structure (copula)
of the variables, rather than on their individual marginal distributions, enabling
efficient closed-form computation through rank-based transformation. Unlike
histogram-based estimators, GCMI avoids the bin selection problem \cite{Ross2014};
and unlike k-nearest neighbor estimators (KSG, \cite{Kraskov2004}), GCMI does not
require sample sizes sufficient for reliable density estimation in the full joint space; its local non-uniformity correction \cite{Gao2015LNC} adds further cost. It allows for reliable MI estimation for both continuous--continuous
and discrete--continuous variable pairs.
For two random variables $X$ and $Y$, mutual information is defined as:
\begin{equation}
I(X,Y) = H(X) + H(Y) - H(X,Y)
\end{equation}
GCMI first transforms the variables into a Gaussian copula space:
\begin{equation}
\tilde{X} = \Phi^{-1}(F_X(X)), \quad \tilde{Y} = \Phi^{-1}(F_Y(Y))
\end{equation}
where $F_X$ and $F_Y$ are empirical cumulative distribution functions (ECDFs), and
$\Phi^{-1}$ is the inverse CDF of the standard normal distribution.
\textbf{Continuous-Continuous Case}
For two continuous variables, MI is estimated using the correlation matrices
$\mathbf{R}_X$ and $\mathbf{R}_Y$ of the transformed variables and the matrix $\mathbf{R}_{XY}$ of both
together:
\begin{equation}
\label{eq:gcmi_continuous}
I_{GCMI}(X,Y) = \frac{1}{2}\log_2\frac{|\mathbf{R}_X|\,|\mathbf{R}_Y|}{|\mathbf{R}_{XY}|},
\end{equation}
which for one-dimensional $X$ and $Y$ reduces to $-\frac{1}{2}\log_2|\mathbf{R}_{XY}|$.

\textbf{Discrete-Continuous Case}
For a discrete variable $X$ with $K$ distinct values and a continuous variable $Y$,
MI is computed via the entropy decomposition:
\begin{equation}
I_{GCMI}(X,Y) = H(Y) - \sum_{k=1}^{K} p(X=k) \, H(Y | X=k)
\end{equation}
where $H(Y)$ is the differential entropy of $Y$ estimated via GCMI, and each
class-conditional entropy $H(Y|X=k)$ is estimated from the subset of data where
$X = k$ using the Gaussian copula assumption.
Entropies were computed with the analytic bias correction for Gaussian entropy
estimates \cite{gcmi}. For a multidimensional variable, each component was
copula-normalized separately and MI was obtained from the joint covariance of
the transformed components.

\textbf{Type-based representation of behavioral variables.}
Before MI estimation, each behavioral variable was represented according to its
type, so that the tuning typical of that type becomes a monotonic dependence, which
the Gaussian-copula estimator measures: a continuous linear
variable $x$ by $[x, (x-c)^2]$, with $c$ its median (or its mean when the median
coincided with the minimum or maximum of the range, as for speed); a circular
variable $\theta$ by its first two harmonics
$[\cos\theta, \sin\theta, \cos 2\theta, \sin 2\theta]$; and spatial position
$(x, y)$ by its full quadratic expansion
$[x, y, (x-c_x)^2, (y-c_y)^2, (x-c_x)(y-c_y)]$. Discrete variables were used
unchanged. The representation depends only on the variable type, not on neural
data, and permutation testing and disentanglement were applied to these
representations. Results are reported under the name of the original variable.

\textbf{Preprocessing of calcium traces.} Negative values of each calcium fluorescence trace
were set to zero, and uniform noise of amplitude $10^{-8}$ was added to break
ties. Multidimensional behavioral variables
(spatial position) received tie-breaking noise of amplitude $10^{-7}$ in the
same way. MI was estimated on every fifth sample of the neural and behavioral
time series (to about 4 and 6~Hz in the sessions recorded at about 20 and
30 frames per second).

For each neuron-behavioral variable pair, letting $A$ denote neural activity and
$Y$ a behavioral variable, the strength of selectivity was quantified as:
\begin{equation}
S = I(A, Y)
\end{equation}
This convention ($A$ for neural activity) is used throughout the remainder of
Methods.

\subsubsection{Statistical assessment and identification of behaviorally selective
neurons}
\label{sec:statistical_assessment}

To evaluate the significance of mutual information (MI) between calcium activity and
behavior, we compared the observed value to a null distribution generated from
randomly time-shifted data \cite{skaggs1993information}.
Specifically, the behavioral variable was circularly shifted relative to the
neural signal $A(t)$ by a random lag $s_i$, and MI was recalculated:
\begin{equation*}
MI_{\text{shuffle}}^{(i)} = I\bigl(A(t), Y(t - s_i)\bigr),
\end{equation*}
where the shift is circular over the recording and is drawn on the downsampled
time grid. Shifts within 10~s of the optimal delay $\tau^{*}$ of the pair
(Section~\ref{sec:temporal_alignment}; five times the 2-s decay time constant
assumed for the indicator) were excluded, so that surrogates do not retain the
true temporal alignment. This procedure was repeated multiple times (referred to as ``shuffles''
throughout) to construct a null distribution of MI values expected by chance.
Surrogate values were computed at the single shifted lag, without repeating the
delay search. The circular time
shifting preserves the temporal
autocorrelation structure inherent in both neural calcium signals and behavioral
variables while destroying their temporal association \cite{Lancaster2018}.
A recent independent comparison of statistical procedures for autocorrelated
calcium-imaging data \cite{NevjenDunn2024} concludes that a combination of cross-validation and a cyclical-shift
permutation test yields higher power than the other methods evaluated while maintaining proper
error-rate control, supporting this choice of null.
This is
critical because calcium indicators produce signals with slow dynamics (decay time
constants of 1-2 seconds for GCaMP6s) \cite{Chen2013}, and behavioral variables often
show strong autocorrelations (e.g., spatial position during continuous locomotion).
Simple random shuffling would destroy these autocorrelations, creating an
unrealistically low null distribution.

The resulting null distribution was modeled using a zero-inflated gamma distribution,
which explicitly accounts for the probability mass at zero that commonly occurs in
MI null distributions:
\begin{equation*}
P(MI = 0) = \pi, \quad p(MI = x) = (1 - \pi) \cdot f_\Gamma(x; \alpha, \beta) \quad \text{for } x > 0
\end{equation*}
where $\pi$ is the zero-inflation parameter estimated as the fraction of shuffled MI
values at or below $10^{-10}$, and $\alpha$ (shape) and $\beta$ (scale) are estimated by
maximum likelihood from the non-zero shuffled MI values only. The gamma component is a natural model for MI null distributions: under the null hypothesis of independence, the sample mutual information is asymptotically chi-squared distributed \cite{Cover2006, Maffulli2022}, and chi-squared is a special case of the gamma family; gamma distributions have also been used directly to model MI null distributions in finite samples \cite{hutter2002distribution}. Zero-inflation accounts for the point mass at zero produced by rank-based estimators when variables are independent.

Significance of the observed MI was assessed using the survival function of the
fitted model:
\begin{equation*}
p\text{-value} = (1 - \pi) \cdot [1 - F_\Gamma(MI_{\text{observed}}; \hat{\alpha}, \hat{\beta})]
\end{equation*}
where $F_\Gamma$ is the gamma CDF. The zero mass does not contribute to tail
probability for positive observed values. This parametric method provides a smoother
and more reliable estimate of tail probabilities than direct empirical quantiles,
which is particularly important when the number of shuffles is limited.

We employed a two-stage procedure to identify neuron-behavior pairs with
statistically significant information content, balancing computational efficiency
with false positive control:

\textbf{Stage 1: Screening}

For each neuron-behavior pair, we generated 100 random shuffles and retained pairs
where:
\begin{equation*}
MI_{\text{observed}} > \max_{i=1}^{100} MI_{\text{shuffle}}^{(i)}
\end{equation*}
This initial screening excludes non-informative pairs early (by chance it retained 1.1--1.5\% of null pairs at zero delay and 4.2--5.3\% with the delay search used in the analyses; see Calibration of false discoveries below), reducing computational demands.

\textbf{Stage 2: Validation}

For retained pairs, we generated $n_{\text{sh}} = 10{,}000$ additional shuffles,
capped at the number of distinct circular shifts of the downsampled recording.
For the recordings analyzed here the cap applied in every session: after
downsampling, the recordings had 2{,}369 distinct shifts in the sessions recorded at
about 20 frames per second and 3{,}571--3{,}572 in those recorded at about 30, and
each shift was used once, with excluded shifts (see above)
replaced by randomly drawn admissible shifts. Two criteria were applied:

(a) \textbf{Non-parametric criterion}: The observed MI must exceed all but at
most four of the $n_{\text{sh}}$ shuffled values:
\begin{equation*}
\#\{i : MI_{\text{shuffle}}^{(i)} \geq MI_{\text{observed}}\} \leq 4,
\end{equation*}
providing a rank-based guard against distribution-fitting failures in the parametric criterion.

(b) \textbf{Parametric criterion}: The $p$-value derived from the fitted gamma
distribution (described above) must fall below a fixed per-pair threshold:
\begin{equation*}
p\text{-value} = (1 - \pi) \cdot [1 - F_\Gamma(MI_{\text{observed}}; \hat{\alpha}, \hat{\beta})]
< p_{\text{threshold}}, \qquad p_{\text{threshold}} = 0.001.
\end{equation*}
The threshold was applied to each neuron-feature pair without correction for
multiple comparisons. The software also implements Holm-Bonferroni \cite{Holm1979},
Bonferroni and Benjamini-Hochberg corrections across the pairs tested in Stage~2;
these are options, not the default of the analysis reported here. We used a fixed
per-pair threshold because with many behavioral features a family-wise correction
lowers the threshold strongly, and its stringency depends on the number of neurons
recorded in a session, so that the same neuron-feature dependence would be accepted
in a small session and rejected in a large one.

For each pair, the chance level of the MI estimate was taken as the mean MI across
its Stage~2 shuffles, $\overline{MI}_{\text{shuffle}}$, and the excess information
as the observed MI minus this chance level. Despite statistical significance, some
neuron-behavior pairs showed dependencies too weak for meaningful interpretation.
To exclude these cases, significant pairs were additionally required to carry a
fixed amount of information above their chance level:
\begin{equation*}
MI_{\text{observed}} - \overline{MI}_{\text{shuffle}} > MI_{\text{threshold}},
\end{equation*}
so that the criterion is comparable across recordings of different length and
features of different dimensionality. $MI_{\text{threshold}}$ was held constant
across all sessions in a given experiment; for the hippocampal dataset analyzed
here, $MI_{\text{threshold}} = 0.04$ bits. For a bivariate Gaussian dependence,
0.04 bits corresponds to a correlation of 0.23. This threshold was applied together
with the $p$-value criterion above, before disentanglement
(Section~\ref{sec:implementation}).

\textbf{Calibration of false discoveries.} Stage~2 shuffles are drawn at the
delay selected on the real data and do not repeat the delay search, so the
$p$-value alone does not account for the choice among the candidate delays
(Section~\ref{sec:temporal_alignment}). We therefore measured the false-positive rate
of the full procedure, with the settings used for the real data, on recordings in
which the neural-behavioral alignment was destroyed, in three null conditions: (i) all
behavioral variables of each of the 64 sessions circularly shifted by 25--75\% of the
recording length; (ii) calcium activity of one mouse paired with the behavior of
another mouse recorded on the same day at the same frame rate (16 sessions); (iii) synthetic
populations without selectivity generated with DRIADA. With delay search, $p < 0.001$
alone was passed by 5.1, 6.4 and 3.4 pairs per 1000 in the three conditions,
against 1.4, 1.8 and 0.8 per 1000 with the delay fixed at zero. With the
excess-information criterion added, as in the reported counts of selective
neurons and pairs, the rates were 0.45, 0.60 and 0.06 per 1000 pairs (one pair),
below the nominal 1 per 1000 of the $p$-value threshold; by variable type, they
exceeded it for spatial position and for body and head direction in at least one
condition (Extended Data Fig.~\ref{fig:ed_calibration}b). Scaled to the hippocampal
dataset, conditions (i) and (ii) correspond to about 2\% and 3\% of the 9{,}598
neuron-feature pairs that passed both criteria before disentanglement.
The anti-selectivity filter was applied in the calibration as in the analysis of
the real data.

\textbf{Anti-selectivity filter.} After the two-stage test and before
disentanglement, significant pairs in which neural activity was not higher in the
preferred range of the variable were excluded as anti-selective. For each pair we
computed the ratio of mean calcium activity, at the optimal delay, inside versus
outside a binary state, or above versus below the median of a linear continuous
variable; pairs with a ratio not exceeding 1 were excluded. For type-based
representations the ratio was computed on the original variable (for example, the
median split of speed for its quadratic representation). Circular variables and
spatial position were not subject to this filter.

The dual use of non-parametric (rank-based) and parametric criteria provides
robustness: the rank-based criterion guards against distribution fitting failures
common with outliers in biological recordings, while the parametric $p$-value extends
the test below the resolution of a finite number of shuffles.
This conservative approach reduces false positives in exploratory
analyses of high-dimensional neural data.

\paragraph{FFT-accelerated permutation testing.}

The two-stage procedure described above requires evaluating MI at thousands of
circular-shifts per neuron-behavior pair. A naive implementation would recompute
MI independently for each shift, yielding a per-pair cost of $O(n_{\text{sh}}
\cdot n)$ where $n$ is the number of time frames and $n_{\text{sh}}$ the number
of shuffles. For a typical experiment with $N$ neurons, $M$ behavioral features,
$D$ delay steps, and $n_{\text{sh}} = 10{,}000$ shuffles, this becomes
prohibitive.

We exploit the fact that circular time shifts correspond to circular
cross-correlations, which can be computed for \emph{all} $n$ possible shifts
simultaneously via the convolution theorem:
\begin{equation}
\label{eq:fft_xcorr}
C_{AY}[s] = \sum_{i=0}^{n-1} A[i] \, Y[(i-s) \bmod n]
           = \mathcal{F}^{-1}\!\bigl[\mathcal{F}[A] \cdot \mathcal{F}[Y]^{*}\bigr][s]
\end{equation}
where $\mathcal{F}$ denotes the discrete Fourier transform and $^{*}$ complex
conjugation. This reduces the cost from $O(n_{\text{sh}} \cdot n)$ to
$O(n \log n)$ for computing MI at all shifts, after which any specific shift is
an $O(1)$ array lookup. The particular form of MI at each shift depends on the
variable types:

\textit{Continuous--continuous case.}
For copula-normalized Gaussian variables, MI reduces to
$I(A, Y_s) = -\tfrac{1}{2}\log_2(1 - r(s)^2)$ where $r(s)$ is the Pearson
correlation at shift $s$. The correlation for all shifts is obtained from
Eq.~\ref{eq:fft_xcorr} as $r(s) = C_{AY}[s] / [(n-1)\,\sigma_A \sigma_Y]$,
requiring a single pair of real-valued FFTs.

\textit{Continuous--discrete case.}
For the continuous neural signal $A$ and a discrete variable $Y$ with $K$ classes,
class-conditional sums at every shift are circular correlations with indicator
functions:
$\sum_i A[i] \cdot \mathds{1}[Y_{(i-s)} = k] = \mathcal{F}^{-1}[\mathcal{F}[A]
\cdot \mathcal{F}[\mathds{1}_{Y=k}]^{*}][s]$.
From these sums and the analogous sums of $A^2$, class-conditional variances and
Gaussian entropies are computed for all shifts simultaneously, yielding MI via
$I = H(A) - \sum_k p(k)\,H(A | Y\!=\!k)$. This requires $K+2$ forward FFTs
(of $A$, $A^2$ and the $K$ indicator functions) and $2K$ inverse FFTs, independent of the
number of shuffles.

\textit{Discrete--discrete case.}
For two discrete variables $X$ and $Y$, contingency tables for all shifts are
obtained via indicator cross-correlations:
$n_{ij}(s) = \mathcal{F}^{-1}[\mathcal{F}[\mathds{1}_{X=i}] \cdot
\mathcal{F}[\mathds{1}_{Y=j}]^{*}][s]$, requiring $K_X \cdot K_Y$ FFT pairs.

\textit{Multivariate case.}
For $d$-dimensional behavioral features (e.g., 2D spatial position), the joint
covariance matrix separates into shift-invariant within-variable blocks and
shift-dependent cross-covariance blocks computed via $d$ FFTs. MI is then
evaluated from the determinant of the joint covariance, using closed-form
vectorized expressions for $d \leq 3$ and a batched log-determinant for
$4 \leq d \leq 36$; this covers the four-dimensional circular and
five-dimensional spatial representations used here.

INTENSE precomputes MI for all $n$ shifts once per neuron-feature pair and stores
the result in an FFT cache. This cache is reused across delay optimization
(Section~\ref{sec:temporal_alignment}), Stage~1 screening, and Stage~2
validation without recomputation. The total cost is $O(N \cdot M \cdot n \log n)$
for the cache build plus $O(N \cdot M \cdot n_{\text{sh}})$ for shift lookups,
compared to $O(N \cdot M \cdot n_{\text{sh}} \cdot n)$ without acceleration,
which makes Stage~2 validation computationally tractable for large-scale
datasets. For one neuron-feature pair of the synthetic benchmark, the significance test
with the FFT cache (delay search and both stages) ran 130 to 220 times faster on one CPU
core than without it, depending on the feature type, and reached the same decision.

\subsubsection{Temporal alignment and delay optimization}
\label{sec:temporal_alignment}

To compute optimal delays between signals, we tested delays between $-2$ and
$+2$~s on the downsampled time grid, in steps of 0.25~s (17 candidate delays) in the
sessions recorded at about 20 frames per second and of 0.17~s (25 candidate delays)
in those recorded at about 30, and selected the delay $\tau^{*}$ that maximized mutual information
$I\bigl(A(t), Y(t-\tau)\bigr)$ between the calcium signal and the behavioral
variable. Positive delays correspond to neural activity following the behavioral
variable. All
subsequent calculations were then performed as described for the zero-delay case,
with the only difference being that the ``true'' mutual information was defined as the
value computed at the optimal time shift.

GCaMP6s shows $\sim$100\,ms rise time and 1--2\,s decay time, creating systematic delays
between spikes and fluorescence \cite{Chen2013}. Different behavioral variables
may have distinct optimal delays for the same neuron.

The $\pm2$ second search window is justified by: (1) the half-decay time of GCaMP6s
transients (0.5--2 seconds) \cite{Chen2013}; (2) behavioral timescales in rodent
decision-making \cite{Pinto2022}; (3) computational tractability while capturing relevant
timescales.

For each neuron-feature pair, the delay maximizing mutual information was selected
from the candidate grid; statistical significance was then assessed at this fixed
delay using circular-shift permutation testing
(Section~\ref{sec:statistical_assessment}). Optimal delays of all signs, including
negative ones, were retained in all downstream analyses.

\subsubsection{Disentangling multiple neuronal selectivity}
\label{sec:disentangling_selectivity}

When multiple behavioral variables are correlated, apparent neuronal selectivity
for one variable may arise from its association with another \cite{NevjenDunn2024, Kriegeskorte2019}.
Disentanglement was applied to every neuron with at least two accepted
features, that is, pairs that passed the significance test with the anti-selectivity
filter and the excess-information criterion
(Section~\ref{sec:statistical_assessment}), separately for each pair $X$, $Y$ of
its accepted features. Let $A$ denote neuronal activity (as in
Section~\ref{sec:mi_computation}). Each pair was resolved by the first of the
following steps that applied; a feature was removed from the neuron's final set
of selectivities if it lost at least one pair.

\textbf{Step 1: priority rules.} Several behavioral labels in this dataset are
nested in, or nearly duplicate, other labels or variables. For these pairs a fixed
priority rule assigned the association to one variable:
\begin{itemize}
\item body direction over head direction;
\item freezing over rest;
\item run, walk, rest and freezing, each over speed;
\item in the arena with objects, each individual object (objects 1--4) over the
pooled object label, and each individual object as well as the pooled object
label over the center zone.
\end{itemize}
A rule was applied whenever the neuron was selective for both variables of the
pair, irrespective of whether the two behavioral variables were significantly
related; conditional MI and interaction information were not computed for pairs
resolved by a rule. The rules were derived from preliminary disentanglement analyses across our behavioral datasets. They fix the outcome for pairs of nested or nearly duplicate labels, which removes double counting and, for neurons driven by the retained variable, errors of the information-based procedure (Step~4) on such pairs. Two of the rules
concern continuous variables. Most speed-selective neurons were also selective for
run, and the rule attributed their selectivity to run; speed
and run were significantly related in every session. Likewise, a neuron
selective for both head and body direction, which are strongly coupled during free
exploration (\fig{fig:results}f), was reported as selective for body direction.

\textbf{Step 2: spatial zones.} For a neuron selective for both spatial
position and a discrete spatial zone (walls, corners, center, objects 1--4,
pooled objects), the zone was retained only if the neuron's place field lay in
it. An occupancy-normalized activity map was computed from the calcium signal at
the full frame rate of the recording, shifted by the optimal delay of the neuron-zone pair, on a $20 \times 20$
grid spanning the range of the trajectory (Gaussian smoothing of activity and
occupancy, $\sigma = 1.5$ bins). Place fields were the connected regions (8-neighborhood) above
half of the peak of this map (relative to its minimum), and the main field was the
one containing the peak. A bin was assigned to the zone if more than half of its
area lay inside the zone, using a geometric mask of the zones drawn for each
session in arena coordinates. The neuron was labeled a zone cell if the peak of
its main field lay in the zone or in an adjacent bin, and the fields whose peaks
lay there carried more than half of the total field activity. In that case
position and the zone were merged into a single place-zone selectivity, reported
under the zone; otherwise position was retained and the zone label discarded. A
neuron could be assigned to several zones. Merged place-zone selectivities were
not tested further against other variables.

\textbf{Step 3: unrelated behavioral variables.} To determine whether the two
behavioral variables were related, we applied the two-stage significance testing
procedure (Section~\ref{sec:statistical_assessment}) to each pair of behavioral
variables, in their type-based representations, at zero delay and with a per-pair
threshold of $p < 0.01$ without correction for multiple comparisons. For pairs of
behavioral variables, only circular shifts shorter than 2~s were excluded from the null
distribution. If the two
behavioral variables were not significantly related, both selectivities were
retained. For \fig{fig:results}f, the mutual information of each pair of behavioral
variables (GCMI estimate in bits, zero delay) was averaged over the 64 sessions, and
pairs that were significantly related in fewer than half of the sessions were shown
in gray.

\textbf{Step 4: conditional mutual information.} Otherwise, we computed the
conditional mutual information $I(A,X|Y)$ depending on the types of $X$ and
$Y$:

1. \textbf{$X$ and $Y$ continuous}:
\begin{equation*}
I(A,X|Y) = H(A,Y) + H(X,Y) - H(A,X,Y) - H(Y)
\end{equation*}
Entropy terms were estimated via the GCMI method with bias correction, on
rank-normalized data.

2. \textbf{$X$ continuous, $Y$ discrete}:

We computed $I(A,X)$ for each discrete value of $Y$ using:
\begin{equation*}
I(A,X) = H(A) + H(X) - H(A,X)
\end{equation*}
The resulting values were weighted by the empirical probabilities of each value of
$Y$.

3. \textbf{$Y$ continuous, $X$ discrete}:
\begin{equation*}
I(A,X|Y) = H(A|Y) - \sum_{k} p(X=k)\, H(A \mid Y, X=k)
\end{equation*}
Gaussian conditional entropies were computed from the residual covariance of the
neural signal given the conditioning variable, restricted to its well-determined
directions, which keeps the estimate stable when a continuous covariate is nearly
constant within a discrete state.

4. \textbf{$X$ and $Y$ discrete}:
\begin{equation*}
I(A,X|Y) = H(A,Y) + H(X,Y) - H(A,X,Y) - H(Y)
\end{equation*}
$H(A,Y)$ was computed separately for each value of $Y$ and averaged; $H(X,Y)$ and
$H(Y)$ were computed using standard discrete entropy formulas, while $H(A,X,Y)$ was
estimated via the sum of conditional entropies across all $X$, $Y$ combinations.

In all four cases each behavioral variable was aligned with the neural signal at
its optimal delay for the neuron, and $I(A,X)$ and $I(A,Y)$ were the observed MI
values of the two pairs at these delays. Small negative estimates of conditional MI
were set to zero.
Note that CMI was not subjected to a separate significance test: significance of the neuron-feature association was already established in the primary MI analysis (Section~\ref{sec:statistical_assessment}). CMI served solely to attribute the significant association by quantifying how much information persists after controlling for a correlated behavioral variable.
We then calculated the interaction information (II) \cite{McGill1954,ghassami2017},
which quantifies redundancy or synergy in the triplet $A$, $X$, and $Y$:
\begin{equation*}
II(A,X,Y) = I(A,X|Y) - I(A,X)
\end{equation*}
This quantity is symmetric with respect to the three variables; we used the
average of the two estimates $I(A,X|Y) - I(A,X)$ and $I(A,Y|X) - I(A,Y)$.
A nonzero $|II|$ indicates that the
three pairwise associations are not independent: negative values arise when
conditioning reduces mutual information (redundancy), while positive values
indicate that conditioning increases it (synergy) \cite{gcmi}. Ghassami and
Kiyavash \cite{ghassami2017} define the same quantity with the opposite sign.

The choice between $X$ and $Y$ was made as follows.

\emph{Redundancy ($II < 0$).} We used the residual fractions, the share of the
information about the neuron that each variable keeps once the other is known,
$r_X = I(A,X|Y)/I(A,X)$ and $r_Y = I(A,Y|X)/I(A,Y)$: a variable with a residual fraction below 0.1 was removed if
the residual fraction of the other variable was at least 0.1; if both were below 0.1, the
variable with the lower MI was removed. Otherwise both variables were retained.

\emph{Synergy ($II \geq 0$).} Only $X$ was retained if $I(A,X)$ exceeded
$I(A,Y)$ more than twofold (or $I(A,Y)$ was below $10^{-6}$ bits) and
$I(A,X|Y) > I(A,Y|X)$; the symmetric condition retained only $Y$. Otherwise both
variables were retained.

When both variables were retained, both selectivities were reported.

\subsubsection{Statistical analysis of delay and selectivity distributions}
\label{sec:statistical_analysis_delays}

For each mouse-session combination, we computed the percentage of recorded
neurons selective to exactly one, exactly two, and three or more variables, and
the number of significant neuron-feature pairs with negative, near-zero and
positive optimal delays as a percentage of recorded neurons. Delays were binned
as negative ($\tau^{*} < -0.75$~s), near-zero ($|\tau^{*}| \leq 0.75$~s) and
positive ($\tau^{*} > 0.75$~s). Categories were compared pairwise with the
two-sided Wilcoxon signed-rank test across mouse-session combinations (SciPy),
without correction for the number of comparisons.
Significance levels are indicated by asterisks:
$*$\,$p < 0.05$, $**$\,$p < 0.01$, $***$\,$p < 0.001$, $****$\,$p < 0.0001$;
non-significant pairs are omitted.
Bars show mean $\pm$ s.e.m. across mouse-session combinations. The exact $p$-values in
\fig{fig:results}c, listed with the test statistics in Supplementary Table~7, were $3.5 \times 10^{-12}$ (one versus two and one versus three or
more variables), $5.2 \times 10^{-12}$ (two versus three or more), $3.9 \times 10^{-12}$
(negative versus near-zero), $3.7 \times 10^{-12}$ (negative versus positive) and
$2.2 \times 10^{-4}$ (near-zero versus positive). Across the 16 mice, each represented by the mean of its sessions, the comparisons of
\fig{fig:results}c held in the same direction as across sessions (two-sided Wilcoxon
signed-rank test over mice). In every mouse, negative delays were the least frequent category and the percentage
of neurons fell from one to two to three or more variables; with all 16 mice differing in
the same direction, these five comparisons reached $p = 3.1 \times 10^{-5}$, the smallest
value the test can give with 16 mice. Near-zero delays exceeded positive delays with
$p = 0.044$.

The share of selective CA1 neurons that remained multiply selective after disentanglement was
compared with the share expected if every neuron coded a single variable by a parametric
bootstrap over synthetic neurons and recording sessions (Supplementary Section~S2).

\subsection{Place cells analysis}
\label{sec:place_cells_analysis}

\subsubsection{Classic place cells analysis}
\label{sec:classic_pc_identification}
Place cells were identified with a Python re-implementation of a previously
described routine \cite{Plusnin2021}. The arena was divided into $4 \times 4$~cm square bins. For
each neuron with at least five calcium events, a spatial event-rate map was
constructed by dividing the number of event onsets by the occupancy time in each
bin, both smoothed with a Gaussian kernel with a standard deviation of one bin; bins the
animal never visited were excluded. Spatial information content was calculated for each
neuron \cite{skaggs1993information}. To assess the significance of spatial
tuning, 1000 shuffled trials were generated by circularly shifting each neuron's
event onsets relative to the animal's trajectory by a random offset of at least 10~s. Spatial
information was recomputed for each shuffled trial, and the observed information
was expressed as a $z$-score, $z = (SI - \mu_{\text{shuffle}})/\sigma_{\text{shuffle}}$,
where $\mu_{\text{shuffle}}$ and $\sigma_{\text{shuffle}}$ are the mean and
standard deviation of the shuffled distribution. Cells with $z > 3$ were
considered place cells. We used $z > 3$
(8{,}246 of 30{,}105 neuron-sessions) rather than the $z > 4$ of the original event-based
analysis of these recordings, because its one-sided normal tail ($p = 0.0013$) is close to
the per-pair threshold $p < 0.001$ of INTENSE, so that both methods label cells at
comparable significance. Where a
$p$-value was needed, it was taken as the one-sided normal tail, $p = 1 - \Phi(z)$. Neurons with fewer than five calcium
events were assigned $z = 0$ and counted as non-place cells.

\subsubsection{Identification of place cells using INTENSE}
\label{sec:intense_pc_identification}
To assess spatial selectivity using the INTENSE framework, we treated the animal's
joint spatial coordinates $C = (x, y)$ as a single behavioral variable,
represented by its five-dimensional quadratic expansion
(Section~\ref{sec:mi_computation}). All information-theoretic measures were
computed for this joint representation, including mutual information
$I(A, C)$. During the shuffling procedure, all components of $C$ were shifted
jointly relative to the neural signal by the same temporal offset, preserving
their coordinate structure.

\subsubsection{Comparison of PC populations}
\label{sec:comparison_pc_populations}
INTENSE place cells were neurons for which spatial position (five-dimensional
representation) passed Stage~2 with $p < 0.001$ and excess MI above 0.04~bits,
before disentanglement; classic place cells were those with spatial-information
$z > 3$ against circularly shifted event trains, computed from calcium events
detected via continuous wavelet transform \cite{Neugornet2021}. Labels and
statistics of all 64 recording sessions were concatenated, and the overlap of the
two sets among all 30{,}105 neuron-sessions was summarized by the Jaccard index,
the number of neuron-sessions in both sets divided by the number in either. Conditional confirmation was computed cumulatively:
for each significance level of one method (INTENSE $p$, using the Stage~2 $p$
where available and otherwise the Stage~1 $p$ reconstructed from the 100 Stage~1
shuffles; classic $p = 1 - \Phi(z)$), the fraction of cells at or above that level
that were place cells under the other method. For rank agreement, every cell was
scored by a continuous statistic of each method (INTENSE: $z$ of the observed MI
against its 100 Stage~1 circular shifts; classic: $z$), and the Jaccard index of
the top-ranked fraction of cells of the two methods was computed for fractions
from 0.5\% to 50\%. To show the overlap expected if the two methods were related only through their
overall rank correlation, the same quantity was computed for pairs of bivariate
normal statistics with the observed Spearman correlation $\rho$ of the two scores
(20 draws); their Pearson correlation was set to $r = 2\sin(\pi\rho/6)$, the value
that gives Spearman correlation $\rho$ for a bivariate normal distribution. Activity-map similarity
(\fig{fig:pc}e) was averaged over cells with $p$ below a common threshold under
both methods, for thresholds from $10^{-1}$ to $10^{-7}$ ($n = 13{,}191$ cells at
$p < 0.1$, $3{,}370$ at $p < 10^{-3}$, $1{,}036$ at $p < 10^{-5}$ and $347$ at
$p < 10^{-7}$).

\subsubsection{Comparison of spatial selectivity}
\label{sec:comparison_spatial_selectivity}

We compared spatial activity patterns between INTENSE and classical methods
to assess methodological convergence.

The comparison of spatial activity patterns required constructing place field maps
from both calcium fluorescence and calcium events. We divided the behavioral
arena into an $8 \times 8$ grid of spatial bins, a resolution that balances spatial
detail with adequate sampling per bin given typical recording durations. For each
spatial bin, we calculated the average neural activity when the animal occupied that
location, weighted by occupancy time to account for non-uniform spatial sampling
common in freely behaving experiments.

For calcium-based activity maps, we first applied a threshold at the 95th percentile
of the fluorescence trace to isolate periods of elevated activity. This percentile
threshold adapts to the dynamic range of each neuron while focusing on
suprathreshold responses most likely to represent action potential-related calcium
transients. The thresholded signal was then averaged within each spatial bin during
occupancy periods, and the resulting map was smoothed with a Gaussian kernel with a
standard deviation of one bin. Event-based activity maps were constructed similarly from the binary event trains of
the wavelet-based detection (Section~\ref{sec:calcium_event_detection}), in which each
event covers its interval.

To quantify the similarity between calcium and event-based place fields, we employed
the Structural Similarity Index (SSIM), a perceptually-motivated metric originally
developed for image quality assessment \cite{Wang2004ssim}. Both activity maps were
normalized to the range [0, 1] before comparison. The SSIM combines information
about luminance, contrast, and structural similarity, making it more robust than
simple pixel-wise comparisons for capturing the overall correspondence of spatial
patterns. We computed SSIM with a data range of 1.0, appropriate for normalized
maps.

Additionally, we calculated complementary metrics to provide a comprehensive
assessment of place field similarity. The mean squared error (MSE) quantified the
average squared difference between normalized activity maps, providing a
straightforward measure of pixel-wise deviation. The peak signal-to-noise ratio
(PSNR) expressed this difference relative to the maximum possible signal, offering a
scale-invariant metric.

\subsection{Synthetic benchmark}
\label{sec:synthetic_benchmark}

\subsubsection{Feature generation}
\label{sec:synthetic_data_generation}

To systematically evaluate INTENSE performance and compare it with existing methods,
we generated synthetic neural datasets with known ground truth connectivity between
neurons and behavioral features, with the generator of DRIADA 1.2.2. Continuous and
discrete features were evaluated in two separate benchmarks, each in five
independent realizations; all conditions of one realization shared the features, the
receptive fields and the random draws of the event generator, so that they differed
only in SNR and $p_{\text{skip}}$.

\textbf{Discrete features:} Binary time series representing discrete behavioral
states (e.g., rearing, object interaction) were generated with controlled temporal
structure. For each feature, we created islands of activity (1s) interspersed with
inactive periods (0s) using:
\begin{itemize}
\item At most 10 active periods per feature
\item Active period durations drawn from a normal distribution with mean 5 seconds
(100 frames at 20 Hz) and $\sigma = \text{mean}/3$
\item Exponentially distributed gaps between active periods
\end{itemize}
These values were chosen so that each feature had enough active periods within a
recording; all methods were compared on the same data.

\textbf{Continuous features:} Continuous behavioral variables (e.g., speed, head
direction analogs) were generated using fractional Brownian motion (fBM) with:
\begin{itemize}
\item Hurst parameter $H = 0.3$ (sub-diffusive, anti-persistent behavior)
\item Length matching experiment duration (2400 seconds)
\item Davies-Harte method for computationally efficient generation
\item Independent generation of each feature
\end{itemize}

\subsubsection{Signal generation}
\label{sec:signal_generation}

\textbf{Feature-neuron coupling:} Each synthetic neuron was randomly assigned to
exactly one behavioral feature, creating a sparse ground truth connectivity matrix.
For continuous features, the neuron's ``receptive field'' was a quantile window
of the feature containing 15\% of its samples, starting at a randomly chosen
quantile.

\textbf{Event generation with controlled reliability:} Neural events were generated
using a two-state inhomogeneous Poisson process \cite{DayanAbbott2001}:
\begin{itemize}
\item Baseline rate: $r_0 = 0.1$ Hz during inactive periods and outside the receptive field
\item Active rate: $r_1 = r_0 \times \text{SNR}$ during active periods and inside the receptive field
\item Skip probability: $p_{\text{skip}} \in [0, 0.8]$ to model response variability
\item Event counts drawn per imaging frame (20 Hz); a frame with one or more events
received a single event
\end{itemize}

The skip probability parameter $p_{\text{skip}}$ controls neuronal response
reliability by randomly deleting active periods with probability $p_{\text{skip}}$,
simulating the stochastic nature of neural responses observed in vivo
\cite{Grijseels2021}.

\textbf{Calcium dynamics:} Discrete spiking events were convolved with a
single-exponential calcium kernel with instantaneous rise to generate
fluorescence traces:
\begin{equation}
A(t) = \sum_i a_i
\exp\!\left(-\frac{t - t_i}{\tau_d}\right) \Theta(t - t_i) + \eta(t)
\end{equation}
where:
\begin{itemize}
\item $t_i$ are event times from the Poisson process
\item $a_i \sim \mathcal{U}(0.5, 2.0)$ are event amplitudes in $\Delta F/F$ units
\item $\tau_d = 2.0$ seconds (decay time constant), matching GCaMP6s kinetics \cite{Chen2013}
\item $\Theta$ is the Heaviside step function
\item $\eta(t) \sim \mathcal{N}(0, 0.1^2)$ is additive Gaussian noise, drawn
independently for each frame
\end{itemize}

\textbf{Noise levels:} To evaluate robustness, we varied the signal-to-noise ratio
(SNR), which is defined as the ratio of firing rates inside versus outside the
neuron's target regions:

$\text{SNR} = \frac{r_1}{r_0}$

where:
$r_0$ = baseline firing rate outside the region of interest (0.1 Hz),
$r_1$ = firing rate within the region of interest ($r_1 = r_0 \times \text{SNR}$)

For example, an SNR of 8 means neurons fire at 0.8 Hz when the behavioral variable
is within their receptive field, compared to 0.1 Hz baseline firing. This definition
allows us to systematically vary the strength of neural selectivity from weak
($\text{SNR} = 2$, 0.2 Hz peak firing) to very strong ($\text{SNR} = 64$, 6.4 Hz peak firing
rate).

We used SNR values spanning $\{2, 4, 8, 16, 32, 64\}$ to cover the range
from heavily corrupted to nearly noiseless conditions.

\subsubsection{Validation dataset composition}
\label{sec:validation_dataset}

Each synthetic dataset comprised:
\begin{itemize}
\item 500 neurons, all selective to features of one type (continuous or discrete,
in separate benchmarks)
\item 20 behavioral features of that type
\item Ground truth connectivity: 5\% (500 true connections out of 10{,}000 possible
pairs)
\item Recording duration: 2400 seconds (48{,}000 frames at 20 Hz)
\item Parameter grid: 6 SNR values $\times$ 9 $p_{\text{skip}}$ values = 54
conditions
\end{itemize}

\subsubsection{Performance evaluation metrics}
\label{sec:performance_metrics}

To quantify detection performance, we computed standard binary classification
metrics by comparing detected significant neuron-feature pairs against ground truth:

\textbf{Precision:} Fraction of detected connections that are true positives:
\begin{equation}
\text{Precision} = \frac{\text{TP}}{\text{TP} + \text{FP}}
\end{equation}

\textbf{Recall:} Fraction of true connections that were detected:
\begin{equation}
\text{Recall} = \frac{\text{TP}}{\text{TP} + \text{FN}}
\end{equation}

\textbf{$F_1$-score:} Harmonic mean of precision and recall:
\begin{equation}
F_1 = 2 \cdot \frac{\text{Precision} \times \text{Recall}}{\text{Precision} +
\text{Recall}}
\end{equation}

where TP = true positives, FP = false positives, and FN = false negatives.

\textbf{Random baseline:} Given the 5\% ground truth connectivity, a random guesser
would achieve:
\begin{equation}
\text{Precision}_{\text{random}} = 0.05
\end{equation}

\subsubsection{Comparison methods}
\label{sec:comparison_methods}

We compared INTENSE against three categories of approaches for neuron-behavior
association:

\textbf{1. Simple criterion methods (without shuffles):}
\begin{itemize}
\item \textbf{For continuous features:} Pearson correlation coefficient with
significance determined via the two-sided Student's t-test. For samples of size
$n$, the test statistic:
\begin{equation}
t = r\sqrt{\frac{n-2}{1-r^2}}
\end{equation}
follows a t-distribution with $n-2$ degrees of freedom under the null hypothesis of
zero correlation \cite{Fisher1915}.

\item \textbf{For discrete features:} Welch's two-sided t-test comparing mean
calcium activity between active/inactive behavioral periods. This test accounts
for unequal variances between groups and uses the Welch-Satterthwaite
approximation for degrees of freedom \cite{Welch1947}:
\begin{equation}
\nu = \frac{\left(\frac{s_1^2}{n_1} +
\frac{s_2^2}{n_2}\right)^2}{\frac{s_1^4}{n_1^2(n_1-1)} + \frac{s_2^4}{n_2^2(n_2-1)}}
\end{equation}
\end{itemize}

\textbf{2. MI-based methods (without shuffles):}
\begin{itemize}
\item \textbf{For discrete features:} we employed the plug-in estimator with
adaptive binning. The number of bins was selected using Sturges' rule extension:
$n_{\text{bins}} = \lceil \log_2(n) \rceil + 1$ \cite{Sturges1926}, with
quantile-based binning to ensure equal sample sizes per bin.

Statistical significance was assessed using the log-likelihood ratio test (G-test
for independence) \cite{Sokal1995}:

\begin{equation}
G = 2 \sum_{i,j} O_{ij} \ln\left(\frac{O_{ij}}{E_{ij}}\right)
\end{equation}

where $O_{ij}$ are observed frequencies and $E_{ij}$ are expected frequencies under
independence. The test statistic follows a $\chi^2$ distribution with $(r-1)(c-1)$
degrees of freedom for $r$ rows and $c$ columns in the contingency table.

\item \textbf{For continuous features:} we converted the GCMI estimate to an
effective correlation \cite{Cover2006}:
\begin{equation}
\rho_{\text{eff}} = \operatorname{sign}(\rho_{\text{sample}}) \cdot \sqrt{1 - 2^{-2 \cdot
\text{MI}}}
\end{equation}
with MI in bits,
and computed a two-sided $p$-value via the t-test for correlation.
\end{itemize}

For all four methods without shuffles, pairs with $p < 0.001$ were counted as detected,
without correction for multiple comparisons: the same per-pair threshold as for the
shuffle-based methods.

\textbf{3. INTENSE-like pipelines:}
\begin{itemize}
\item \textbf{correlation-based:} INTENSE pipeline using vectorized Pearson
correlation as the similarity metric. While maintaining the full statistical rigor
of the two-stage procedure, this test addresses the linear dependency between the
signals.

\item \textbf{average-based:} INTENSE pipeline for discrete features with the ratio
of mean calcium activity inside versus outside the behavioral state as the
similarity metric.

\item \textbf{MI-based (INTENSE):} Full INTENSE pipeline with Gaussian Copula Mutual
Information (GCMI) \cite{gcmi}. GCMI transforms variables to a Gaussian copula
representation where MI has the closed form of Eq.~\ref{eq:gcmi_continuous}.
This approach is robust to outliers and requires no binning parameters.
\end{itemize}

All INTENSE-like methods employed the two-stage testing procedure described in
Section~\ref{sec:statistical_assessment} with the settings of the analysis of the
recordings: 100 Stage~1 and 10{,}000 Stage~2 shuffles (capped at the 9{,}600
circular shifts of the downsampled 2400-s recordings), $p < 0.001$ per pair without
correction for multiple comparisons, delay search within $\pm 2$~s and the type-based
representation of the variables for MI. $p$-values came from a
zero-inflated gamma distribution fitted to the shuffled values for MI and from a
normal distribution for correlation and the activity ratio. The excess-information
criterion of 0.04~bits, which has no counterpart for correlation or the activity
ratio, was not applied in the synthetic benchmark, so that the three shuffle-based
methods were compared by the same significance test. The anti-selectivity filter was
not applied either: it compares activity above and below the median of a continuous
variable and would by construction remove the synthetic neurons tuned to the lower
half of their variable.

\subsubsection{Experimental protocol}
\label{sec:experimental_protocol}

For each parameter combination (SNR, $p_{\text{skip}}$), we performed 5 independent
random initializations and averaged the resulting precision and recall; $F_1$ was
computed from the averaged precision and recall.
This averaging reduces the impact of random fluctuations and provides more stable
performance estimates. For the two-stage analysis (Section~\ref{sec:two_stage_testing}),
precision was also computed for the Stage~1 candidates.
The comprehensive evaluation covered:
\begin{itemize}
\item 6 SNR values: $\{2, 4, 8, 16, 32, 64\}$
\item 9 $p_{\text{skip}}$ values: $\{0.0, 0.1, 0.2, 0.3, 0.4, 0.5, 0.6, 0.7, 0.8\}$
\item Total: $6 \times 9 \times 5 = 270$ synthetic datasets per feature type
\end{itemize}

Downsampling by factor 5 (from 20 to 4~Hz for the synthetic data, to about 4 or 6~Hz for the recordings) was applied to both synthetic and experimental data across all methods to reduce computational load while maintaining sufficient temporal resolution to detect neuron-feature associations, given calcium transient decay times of 1--2 seconds for GCaMP6s \cite{Chen2013}.

\subsection{AI usage disclosure}
AI coding assistants (Claude, Anthropic) were used to search for errors in the
software and to test it, to write and check analysis and figure scripts, to check
numbers and references, and for language editing of the manuscript. The authors
reviewed all outputs of these tools and take full responsibility for the content
of this article.

\clearpage
\section*{Extended Data}
\begingroup
\renewcommand{\figurename}{Extended Data Fig.}
\renewcommand{\thefigure}{\arabic{figure}}
\renewcommand{\theHfigure}{ED\arabic{figure}}
\setcounter{figure}{0}
\begin{figure}[htbp]
  \centering
  \includegraphics[width=\linewidth]{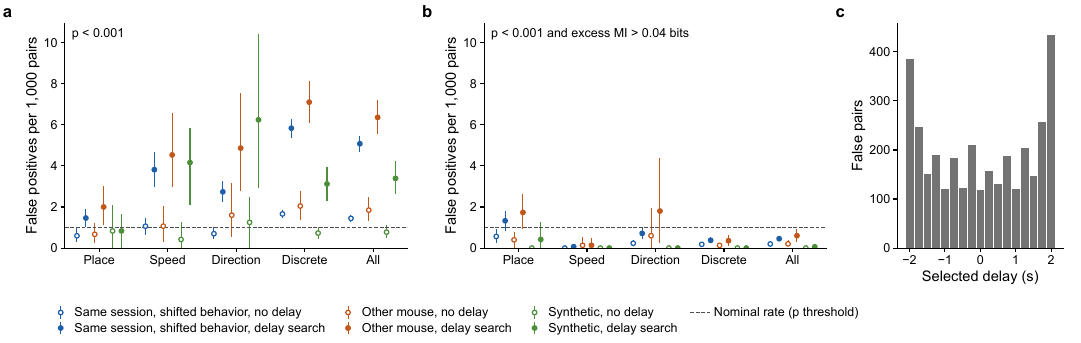}
  \caption{\textbf{Calibration of false discoveries on recordings without
neural--behavioral alignment.} \textbf{a}, Neuron--feature pairs passing Stage~2
with $p < 0.001$, per 1,000 pairs tested, by variable type (place, spatial
position; speed; direction, body and head direction, or head direction alone in the
synthetic populations; discrete, all discrete variables;
all, all features), in three null conditions: behavioral variables of the same
session circularly shifted by 25--75\% of the recording (blue), calcium activity of
one mouse paired with the behavior of another mouse recorded on the same day
(orange) and synthetic populations without selectivity (green) (see Methods:
Section~\ref{sec:statistical_assessment}). Open circles, delay fixed at zero; filled circles, delay search within
$\pm 2$~s, as in the analysis of the real data. Error bars, 95\% bootstrap
intervals over sessions (over recordings for the synthetic condition). Dashed
line, nominal rate of the per-pair threshold (1 per 1,000). \textbf{b}, Same with
the excess-information criterion ($> 0.04$~bits) added, as in the reported
counts of selective neurons and pairs. \textbf{c}, Optimal delays selected for the false pairs of
\textbf{a} found with the delay search, all three null conditions pooled.}
  \label{fig:ed_calibration}
\end{figure}
\endgroup
\clearpage

\section*{Data availability}
\label{sec:data_availability}
The hippocampal calcium imaging dataset analyzed in this study (16 mice, 64
sessions of the open field enriched with objects, labeled EOF in the dataset; fluorescence traces, behavioral
variables and session metadata in NWB format, with the raw miniscope and behavioral videos) is available in the DANDI
Archive as part of Dandiset 002047
(\href{https://dandiarchive.org/dandiset/002047}{https://dandiarchive.org/dandiset/002047}). The synthetic datasets used for
calibration and validation are generated by the scripts listed under Code
availability.
The medial entorhinal cortex dataset analyzed in this study was published by its authors \cite{Obenhaus2022}
and is available from the NIRD Research Data Archive
(\href{https://doi.org/10.11582/2022.00005}{https://doi.org/10.11582/2022.00005}).
Source data are provided with this paper.

\section*{Code availability}
\label{sec:code_availability}
INTENSE is part of the open-source DRIADA (Dimensionality Reduction for
Integrated Activity Data Analysis) Python package \cite{driada}, released under
the MIT License. All analyses used DRIADA version 1.2.2, available on GitHub
(\href{https://github.com/iabs-neuro/driada}{https://github.com/iabs-neuro/driada}),
on PyPI (\texttt{pip install driada==1.2.2}) and archived on Zenodo
(\href{https://doi.org/10.5281/zenodo.23986480}{doi:10.5281/zenodo.23986480}); documentation is
available at \href{https://driada.readthedocs.io}{https://driada.readthedocs.io}.
The scripts that reproduce the analyses, numbers and figures of this paper are in
the \texttt{publications/intense} directory of the DRIADA repository; its README
describes how to run them and lists the settings that decide the numbers, including
the random seed. With the same seed, numbers obtained on synthetic data are reproduced
from run to run up to differences in the last digit.
INTENSE builds on the Gaussian Copula Mutual Information (GCMI) framework
\cite{gcmi}; the reference GCMI implementation is available at
\href{https://github.com/robince/gcmi}{https://github.com/robince/gcmi}.
Miniscope videos were processed with the BEARMIND pipeline
(\href{https://github.com/iabs-neuro/bearmind}{https://github.com/iabs-neuro/bearmind})
and behavior was analyzed with the Sphynx pipeline \cite{Plusnin1}
(\href{https://github.com/iabs-neuro/sphynx}{https://github.com/iabs-neuro/sphynx}); neural and behavioral time series were
aligned with the IABS data synchronizer
(\href{https://github.com/iabs-neuro/iabs_synchronizer}{https://github.com/iabs-neuro/iabs\_synchronizer}).
Continuous wavelet transforms with generalized Morse wavelets were computed with
the Python package ssqueezepy \cite{ssqueezepy}.

\section*{Author contributions}
\label{sec:author_contributions}
\textbf{Nikita Pospelov:} Conceptualization, Software, Methodology, Validation, Formal analysis,
Data curation, Writing -- original draft, Writing -- review \& editing, Visualization, Project
administration. \textbf{Olga Ivashkina:} Methodology, Validation, Investigation, Data curation,
Resources, Writing -- original draft, Writing -- review \& editing, Visualization.
\textbf{Viktor Plusnin:} Software, Validation, Investigation, Data curation, Writing -- review \&
editing. \textbf{Olga Rogozhnikova:} Investigation, Software, Formal analysis, Writing -- review \&
editing. \textbf{Anna Ivanova:} Investigation, Data curation, Writing -- review \& editing.
\textbf{Vladimir Sotskov:} Methodology, Investigation, Resources, Writing -- review \& editing.
\textbf{Ksenia Toropova:} Methodology, Validation, Investigation, Data curation, Writing -- review
\& editing. \textbf{Vladik Avetisov:} Formal analysis, Writing -- review \& editing.
\textbf{Konstantin Anokhin:} Conceptualization, Supervision, Resources, Funding acquisition,
Writing -- review \& editing.

\section*{Competing interests}
\label{sec:competing_interests}
The authors declare no competing interests.

\section*{Funding}
\label{sec:funding}
This work was supported by the Non-Commercial Foundation for Support of Science and
Education ``INTELLECT''.

\section*{Acknowledgments}
\label{sec:acknowledgements}
We are grateful to Sergey Nechaev and Olga Martynova for valuable comments and
discussions, and to Nikita Savelev for help with experiments.

\clearpage
\includepdf[pages=-]{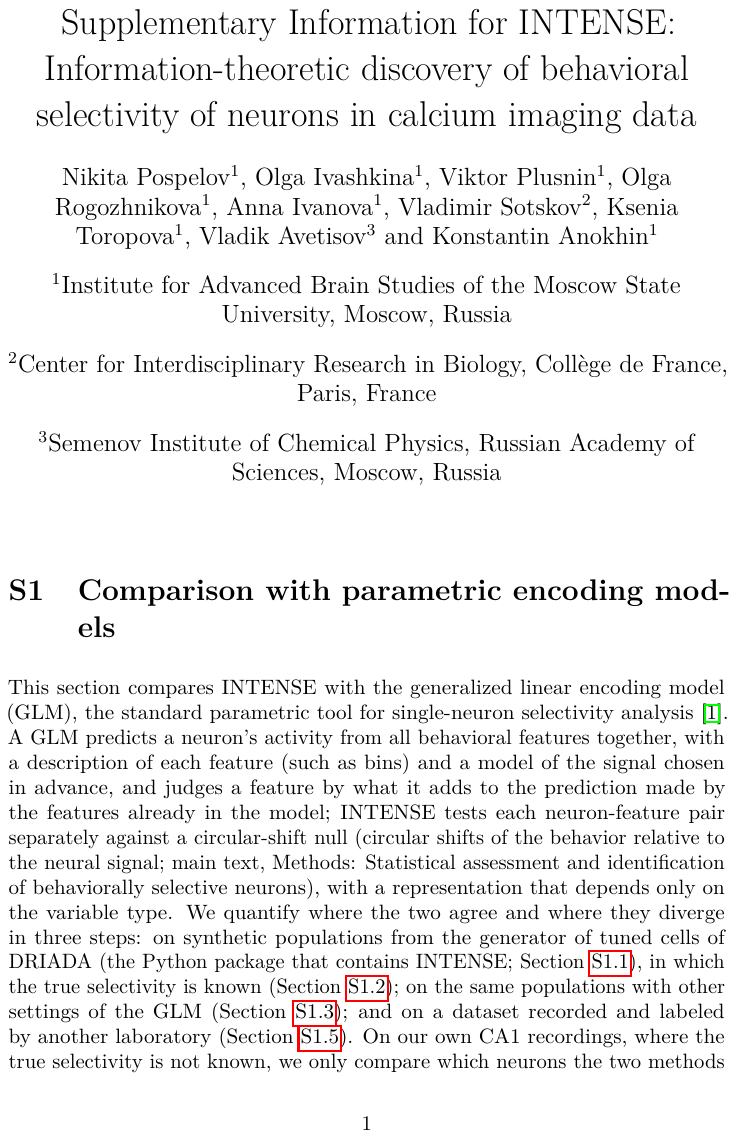}
\end{document}